\documentclass[%
 reprint,
 superscriptaddress,
 showpacs,preprintnumbers,
 amsmath,amssymb,
 aps,
 prc,
 nofootinbib
]{revtex4-1}

\usepackage{bm}
\usepackage{graphicx,color}
\usepackage{multirow}

\newcommand{\VF}[1]{\ensuremath{{}^V\!\!F^0_{#1}}}
\newcommand{\AF}[1]{\ensuremath{{}^A\!F^0_{#1}}}
\newcommand{\VAF}[1]{\ensuremath{{}^{V,A}\!F^0_{#1}}}

\begin{document}

\title{Measurement of the $\bm{2^+\rightarrow 0^+}$ ground-state transition in the $\bm{\beta}$ decay of $\bm{^{20}}$F}


\author{O.~S.~Kirsebom}
\email[Corresponding author: ]{oliver.kirsebom@dal.ca}
\affiliation{Department of Physics and Astronomy, Aarhus University, DK-8000 Aarhus C, Denmark}
\affiliation{Institute for Big Data Analytics, Dalhousie University, Halifax, NS, B3H 4R2, Canada}


\author{M.~Hukkanen}
\affiliation{University of Jyvaskyla, Department of Physics, P.O.\ Box 35, FI-40014, University of Jyvaskyla, Finland}

\author{A.~Kankainen}
\affiliation{University of Jyvaskyla, Department of Physics, P.O.\ Box 35, FI-40014, University of Jyvaskyla, Finland}

\author{W.~H.~Trzaska}
\affiliation{University of Jyvaskyla, Department of Physics, P.O.\ Box 35, FI-40014, University of Jyvaskyla, Finland}

\author{D.~F.~Str\"omberg}
\affiliation{Institut f{\"u}r Kernphysik
  (Theoriezentrum), Technische Universit{\"a}t Darmstadt,
  Schlossgartenstra{\ss}e 2, 64289 Darmstadt, Germany}
\affiliation{GSI Helmholtzzentrum f\"ur Schwerionenforschung,
  Planckstra{\ss}e~1, 64291 Darmstadt, Germany}

\author{G.~Mart\'inez-Pinedo}
\email{g.martinez@gsi.de}
\affiliation{GSI Helmholtzzentrum f\"ur Schwerionenforschung,
  Planckstra{\ss}e~1, 64291 Darmstadt, Germany}
\affiliation{Institut f{\"u}r Kernphysik
  (Theoriezentrum), Technische Universit{\"a}t Darmstadt,
  Schlossgartenstra{\ss}e 2, 64289 Darmstadt, Germany}


\author{K.~Andersen}
\affiliation{Department of Physics and Astronomy, Aarhus University, DK-8000 Aarhus C, Denmark}

\author{E.~Bodewits}
\affiliation{SCIONIX Holland B.V., Regulierenring 5, 3981 LA Bunnik, The Netherlands}

\author{B.~A.~Brown}
\affiliation{National Superconducting Cyclotron Laboratory, Michigan State University, East Lansing, Michigan 48824, USA}

\author{L.~Canete}
\affiliation{University of Jyvaskyla, Department of Physics, P.O.\ Box 35, FI-40014, University of Jyvaskyla, Finland}

\author{J.~Cederk\"all}
\affiliation{Department of Physics, Lund University, SE-22100 Lund, Sweden}

\author{T.~Enqvist}
\affiliation{University of Oulu, Oulu Southern Institute, FI-90014, Finland}

\author{T.~Eronen}
\affiliation{University of Jyvaskyla, Department of Physics, P.O.\ Box 35, FI-40014, University of Jyvaskyla, Finland}

\author{H.~O.~U.~Fynbo}
\affiliation{Department of Physics and Astronomy, Aarhus University, DK-8000 Aarhus C, Denmark}

\author{S.~Geldhof}
\affiliation{University of Jyvaskyla, Department of Physics, P.O.\ Box 35, FI-40014, University of Jyvaskyla, Finland}

\author{R.~de Groote}
\affiliation{University of Jyvaskyla, Department of Physics, P.O.\ Box 35, FI-40014, University of Jyvaskyla, Finland}

\author{D.~G.~Jenkins}
\affiliation{Department of Physics, University of York, York YO10 5DD, United Kingdom}

\author{A.~Jokinen}
\affiliation{University of Jyvaskyla, Department of Physics, P.O.\ Box 35, FI-40014, University of Jyvaskyla, Finland}

\author{P.~Joshi}
\affiliation{Department of Physics, University of York, York YO10 5DD, United Kingdom}

\author{A.~Khanam}
\affiliation{University of Jyvaskyla, Department of Physics, P.O.\ Box 35, FI-40014, University of Jyvaskyla, Finland}
\affiliation{Aalto University, P.O. Box 11000, FI-00076 Aalto, Finland}

\author{J.~Kostensalo}  
\affiliation{University of Jyvaskyla, Department of Physics, P.O.\ Box 35, FI-40014, University of Jyvaskyla, Finland}

\author{P.~Kuusiniemi}
\affiliation{University of Oulu, Oulu Southern Institute, FI-90014, Finland}

\author{K.~Langanke}
\affiliation{GSI Helmholtzzentrum f\"ur Schwerionenforschung,
  Planckstra{\ss}e~1, 64291 Darmstadt, Germany}
\affiliation{Institut f{\"u}r Kernphysik
  (Theoriezentrum), Technische Universit{\"a}t Darmstadt,
  Schlossgartenstra{\ss}e 2, 64289 Darmstadt, Germany}

\author{I.~Moore}
\affiliation{University of Jyvaskyla, Department of Physics, P.O.\ Box 35, FI-40014, University of Jyvaskyla, Finland}

\author{M.~Munch}
\affiliation{Department of Physics and Astronomy, Aarhus University, DK-8000 Aarhus C, Denmark}

\author{D.~A.~Nesterenko}
\affiliation{University of Jyvaskyla, Department of Physics, P.O.\ Box 35, FI-40014, University of Jyvaskyla, Finland}

\author{J.~D.~Ovejas}
\affiliation{Instituto de Estructura de la Materia, CSIC, E-28006 Madrid, Spain}

\author{H.~Penttil\"a}
\affiliation{University of Jyvaskyla, Department of Physics, P.O.\ Box 35, FI-40014, University of Jyvaskyla, Finland}

\author{I.~Pohjalainen}
\affiliation{University of Jyvaskyla, Department of Physics, P.O.\ Box 35, FI-40014, University of Jyvaskyla, Finland}

\author{M.~Reponen}
\affiliation{University of Jyvaskyla, Department of Physics, P.O.\ Box 35, FI-40014, University of Jyvaskyla, Finland}

\author{S.~Rinta-Antila}
\affiliation{University of Jyvaskyla, Department of Physics, P.O.\ Box 35, FI-40014, University of Jyvaskyla, Finland}

\author{K.~Riisager}
\affiliation{Department of Physics and Astronomy, Aarhus University, DK-8000 Aarhus C, Denmark}

\author{A.~de Roubin}
\affiliation{University of Jyvaskyla, Department of Physics, P.O.\ Box 35, FI-40014, University of Jyvaskyla, Finland}

\author{P.~Schotanus}
\affiliation{SCIONIX Holland B.V., Regulierenring 5, 3981 LA Bunnik, The Netherlands}

\author{P.~C.~Srivastava}
\affiliation{Department of Physics, Indian Institute of Technology, Roorkee 247667, India}

\author{J.~Suhonen}  
\affiliation{University of Jyvaskyla, Department of Physics, P.O.\ Box 35, FI-40014, University of Jyvaskyla, Finland}

\author{J.~A.~Swartz}
\affiliation{Department of Physics and Astronomy, Aarhus University, DK-8000 Aarhus C, Denmark}

\author{O.~Tengblad}
\affiliation{Instituto de Estructura de la Materia, CSIC, E-28006 Madrid, Spain}

\author{M.~Vilen}
\affiliation{University of Jyvaskyla, Department of Physics, P.O.\ Box 35, FI-40014, University of Jyvaskyla, Finland}

\author{S.~V\'inals} 
\affiliation{Instituto de Estructura de la Materia, CSIC, E-28006 Madrid, Spain}

\author{J.~\"Ayst\"o}
\affiliation{University of Jyvaskyla, Department of Physics, P.O.\ Box 35, FI-40014, University of Jyvaskyla, Finland}

\date{\today}

\begin{abstract}

We report the first detection of the second-forbidden, non-unique, 
$2^+\rightarrow 0^+$, ground-state transition in the 
$\beta$ decay of $^{20}$F. %
A low-energy, mass-separated $^{20}\rm{F}^+$ beam produced at the IGISOL 
facility in Jyv\"askyl\"a, Finland, was implanted in a thin 
carbon foil and the $\beta$ spectrum measured using a 
magnetic transporter and a plastic-scintillator detector. %
The $\beta$-decay branching ratio inferred from the measurement is 
$b_{\beta} = [ 0.41\pm 0.08\textrm{(stat)}\pm 0.07\textrm{(sys)}] \times 10^{-5}$
corresponding to $\log ft = 10.89(11)$, making this one of the strongest 
second-forbidden, non-unique $\beta$ transitions ever measured. %
The experimental result is supported by shell-model calculations 
and has significant implications for the final evolution of stars 
that develop degenerate oxygen-neon cores. %
Using the new experimental data, we argue that the astrophysical 
electron-capture rate on $^{20}$Ne is now known to within better than 
25\% at the relevant temperatures and densities.

\end{abstract}

\maketitle


\section{Introduction}

Second-forbidden, non-unique $\beta$ transitions ($\Delta J = 2$, $\Delta \pi = \textrm{no}$) 
typically have very small branching ratios, which makes their detection rather challenging. 
Only around 27 such transitions have been observed~\cite{singh1998}. Measurements of the 
rates and shapes of forbidden $\beta$ transitions provide insights 
into nuclear structure and occasionally also into astrophysical processes. %

\begin{figure}[htb]
  \centering
  \includegraphics[width=0.95\linewidth]{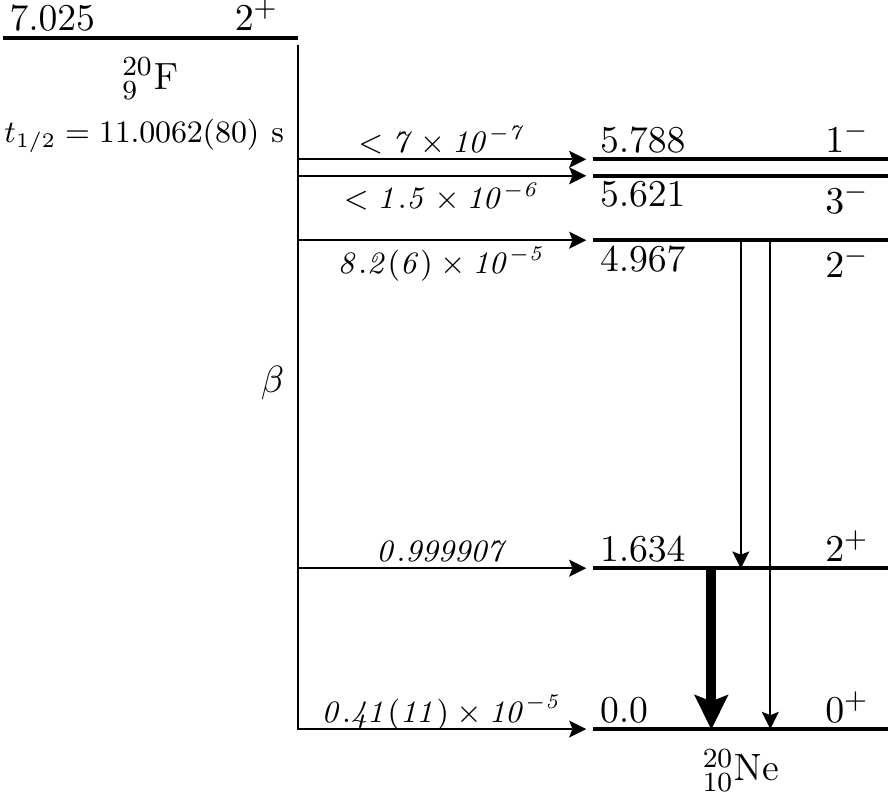}
\caption{\label{fig:decayscheme} $^{20}$F $\beta$-decay scheme~\cite{tunl_A20, wilds2007} 
including the newly observed ground-state transition. The bold arrow indicates the $\gamma$-ray 
transition used for absolute normalisation. The branching ratios of the individual 
$\beta$-decay transitions are shown in italic. 
Energies are in MeV relative to the $^{20}$Ne ground state.}
\end{figure}

Recent studies have highlighted the importance of the second-forbidden, 
non-unique, electron-capture transition from the $0^+$ ground state of $^{20}$Ne 
to the $2^+$ ground state of $^{20}$F for the final evolution of 
stars of 7--11 solar masses that develop degenerate oxygen-neon cores~\cite{pinedo14, schwab2015, schwab2017}. 
The strength of the transition is, however, not well constrained, neither 
experimentally nor theoretically, making an experimental determination 
highly desirable. %
The strength may be determined from the branching ratio of the inverse 
$2^+\rightarrow 0^+$ transition in the $\beta$ decay of $^{20}$F 
(Fig.~\ref{fig:decayscheme}), but this transition is not easily 
detected as it is masked by the much faster, allowed, $2^+\rightarrow 2^+$ 
transition to the first-excited state in $^{20}$Ne. %
Indeed, previous attempts to detect the $2^+\rightarrow 0^+$ transition 
have been unsuccessful~\cite{wong54, glickstein63, glickstein63_erratum, calaprice78} yielding 
a rough upper limit of $\sim 10^{-5}$ on the branching ratio~\cite{calaprice78}. %
The $\beta$-decay endpoint energies for the ground-state and first-excited 
state transitions are 7.025~MeV and 5.391~MeV, respectively\footnote{The endpoint energies are known to sub-keV precision~\cite{tunl_A20, ame2016}.}, 
leaving a rather narrow energy window for the detection of the ground-state 
transition. %
Here, we report the first successful measurement of the second-forbidden, 
non-unique, $2^+\rightarrow 0^+$ transition in the $\beta$ decay of $^{20}$F, 
present shell-model calculations which corroborate the experimental result, 
and determine the impact on the astrophysical electron-capture rate on $^{20}$Ne. 
The astrophysical implications for the evolution of stars of 
7--11 solar masses are dealt with elsewhere~\cite{Kirsebom.Jones.ea:2019}.


\section{Experimental setup}\label{sec:setup}

%
The experiment was performed at the IGISOL-4 facility of the JYFL Accelerator Laboratory 
in Jyv\"askyl\"a, Finland~\cite{arje1985, moore2013}. 
Radioactive ions of $^{12}$B$^+$ and $^{20}$F$^+$ were 
produced via $(d,p)$ reactions on targets of boron (B) and barium 
flouride (BaF$_2$). 
The K130 cyclotron was used to produce the deuteron beam, which 
had an energy of 9~MeV and an intensity of around 10~$\mu$A. 
For the production of the $^{20}$F$^+$ ions, a 53~$\mu$m thick tantalum (Ta) degrader was used to reduce the beam energy to 6~MeV. The B and BaF$_2$ 
targets were 0.5~mg/cm$^2$ and 1.2~mg/cm$^2$ thick with backings 
of 4.5~$\mu$m tantalum (Ta) and 2~$\mu$m tungsten (W), respectively. %
The reaction products were thermalized in the IGISOL ion guide gas cell, using helium at a pressure of around 100~mbar for $^{12}$B and 250~mbar for $^{20}$F, and extracted with a sextupole ion guide \cite{Karvonen2008}. After acceleration to 30~keV, 
the ions were separated based on their mass-to-charge ratio using a 
dipole magnet, before being guided to the experimental station where they were stopped in a thin (50~$\mu$g/cm$^2$) carbon foil. 
The $\beta$ spectrum of $^{20}$F was the primary interest of the experiment 
while the $\beta$ spectrum of $^{12}$B provided important calibration data. %

\begin{figure}[h]
  \includegraphics[width=\linewidth]{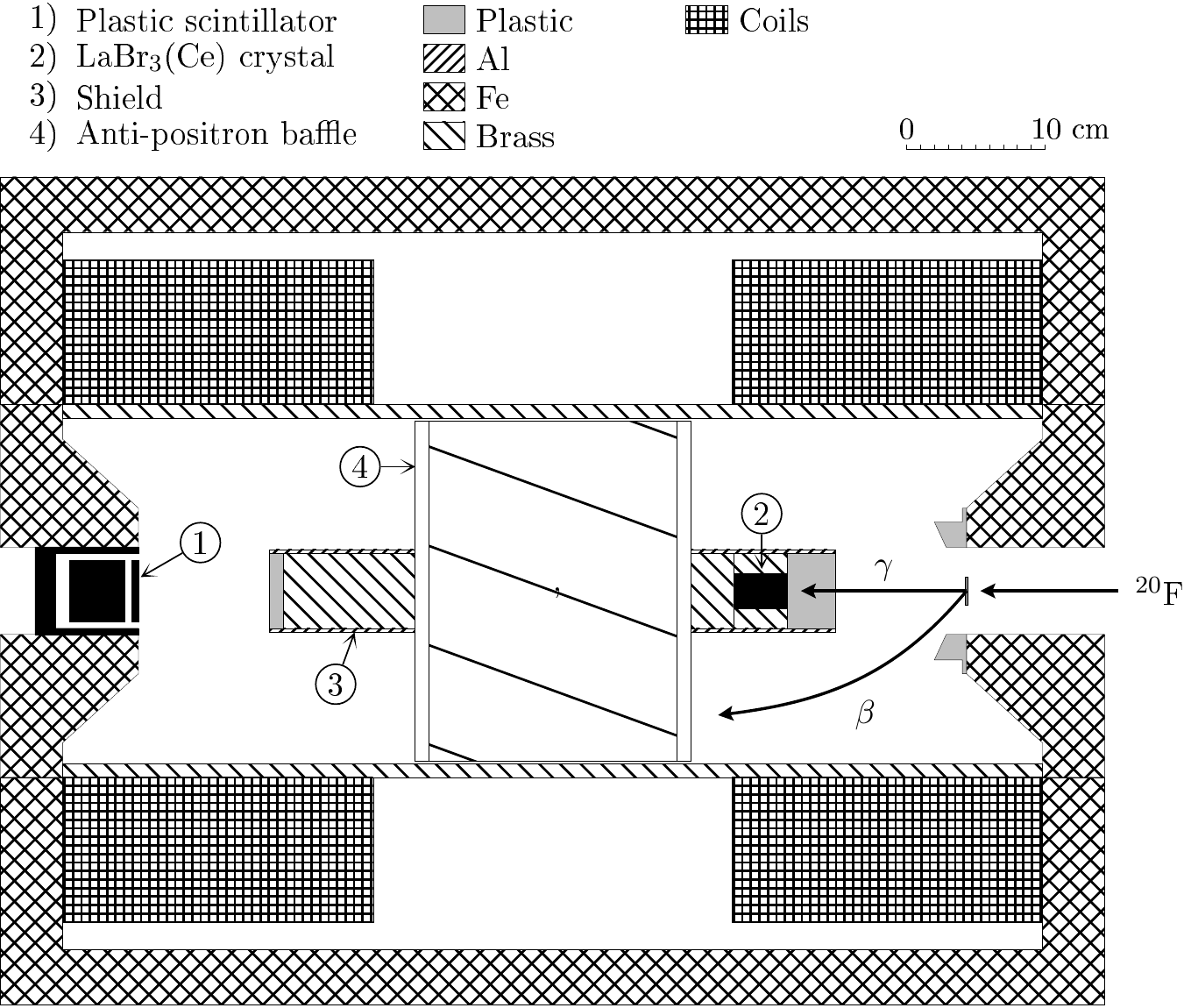} 
 \caption{\label{fig:setup} Schematic diagram of the setup. The
    $^{20}$F beam comes in from the right and is stopped in the
    catcher foil. The 1.63-MeV $\gamma$ ray is detected in the
    LaBr$_3$(Ce) detector (2) that sits behind 3.5 cm of plastic,
    while the electron follows a helical path to the focal plane where
    it is detected in the plastic-scintillator detector (1).}
\end{figure}

The detection system, shown in Fig.~\ref{fig:setup}, consisted of a 
Siegbahn-Sl\"atis type intermediate-image magnetic electron transporter~\cite{siegbahn1949} 
combined with an energy-dispersive detector. %
Such an arrangement is well suited for the measurement of rare ground-state 
transitions in nuclear $\beta$ decays as the effective solid angle of the detector, and hence the 
count rate, is greatly increased by the focussing action of the magnetic field. %
Furthermore, and equally important, the shield on the centre axis prevents $\gamma$ rays 
and electrons produced by transitions to excited states in the daughter nucleus from 
reaching the detector. This essentially 
eliminates $\beta\gamma$ summing and $\beta\beta$ pile-up as sources 
of background and leads to an improved sensitivity towards the ground-state transition. %

\begin{figure}[htb]
  \includegraphics[width=\linewidth]{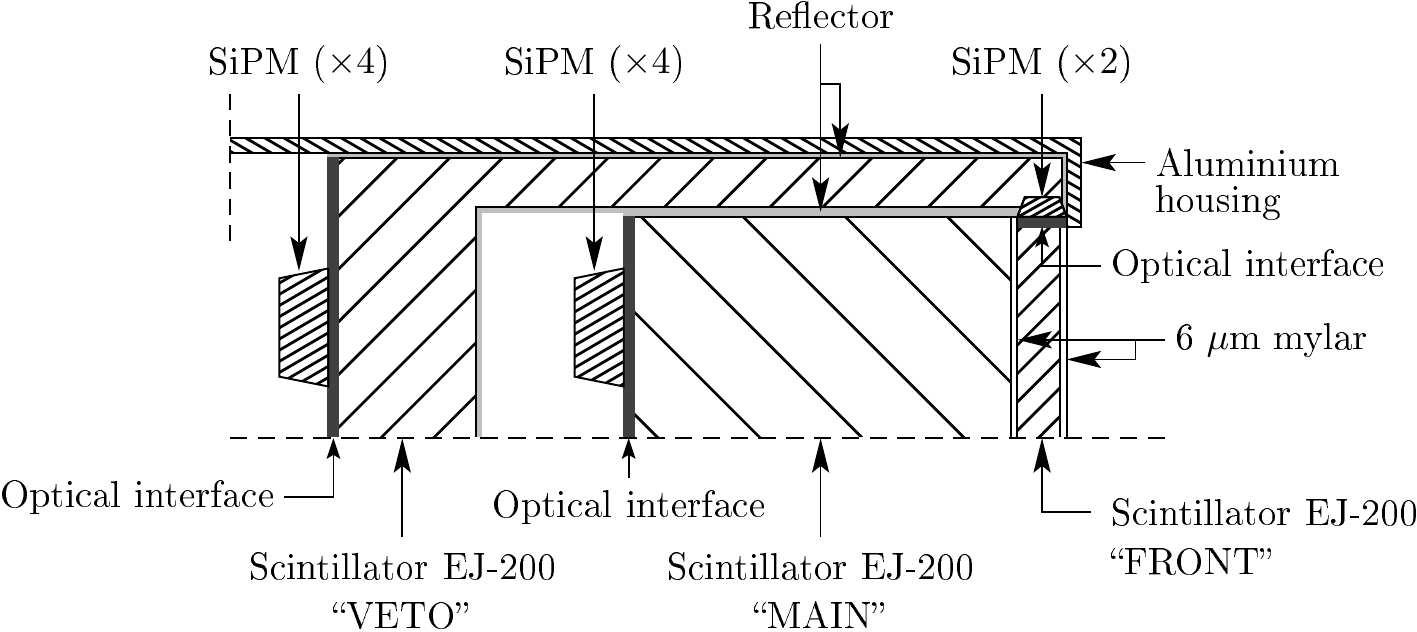}
  \caption{\label{fig:scionix} Schematic diagram of
    plastic-scintillator detector. Built-in amplifiers and cables to
    the SiPMs are not shown. The dimensions
    ($\textrm{diameter}\times\textrm{length}$) of the
    plastic-scintillator volumes are: $45\times 5$~mm$^2$ (FRONT),
    $45\times 40$~mm$^2$ (MAIN) and $55\times 75$~mm$^2$ (VETO). The
    other diameter of the aluminium housing is 60~mm.}
\end{figure}

The magnetic transporter was constructed at the Department of Physics, University of 
Jyv\"askyl\"a (JYFL) in the 1980s~\cite{julin1988}, but has been fully refurbished for 
the present experiment. 
The $\beta$ detector, shown in Fig.~\ref{fig:scionix}, has the shape of a cylinder and consists of a 5-mm thick 
outer detector, used as a veto against cosmic rays, and a 45~mm $\times$ 45~mm 
inner detector, used to measure the full energy of the electrons. 
The inner detector is further subdivided into a 5-mm thick front detector and a 
40-mm thick main detector to provide additional discriminatory power. %
All three detectors (veto, front and main) are plastic scintillators read out with 
silicon photomultipliers. %
The detector dimensions represent a compromise between the requirement to fully stop a 
significant fraction of the most energetic electrons (the nominal range of 
7-MeV electrons in plastic is 35~mm) and the requirement to minimize the cosmic-ray 
exposure. %

Some of the calibration data presented in this paper were obtained using 
an earlier, two-channel version of the three-channel detector that we have just described. The two detectors have identical dimensions and only differ in one respect: the inner volume of the two-channel detector is 
not divided into a front and a main section. 
When necessary we use the labels v1 and v2 to distinguish between the 
two-channel (v1) and three-channel (v2) detector.
When no label is provided it is understood that the three-channel detector 
was used.

A small LaBr$_3$(Ce) crystal placed inside the shield on the centre axis was used 
to detect the 1.63-MeV $\gamma$ ray from the decay of $^{20}$F, 
thereby providing absolute normalisation of the $\beta$ spectrum. %
Finally, a baffle placed at the centre of the magnet prevented positrons, 
which spiral in the opposite direction of electrons, from reaching the detector 
thereby eliminating positron emitters as a potential source of background, 
while reducing the electron flux by only 11\%.

\section{Data analysis and results}\label{sec:results}

\subsection{Characterization of the $\bm{\beta}$ response}

\begin{figure}[htb]
  \includegraphics[width=\linewidth]{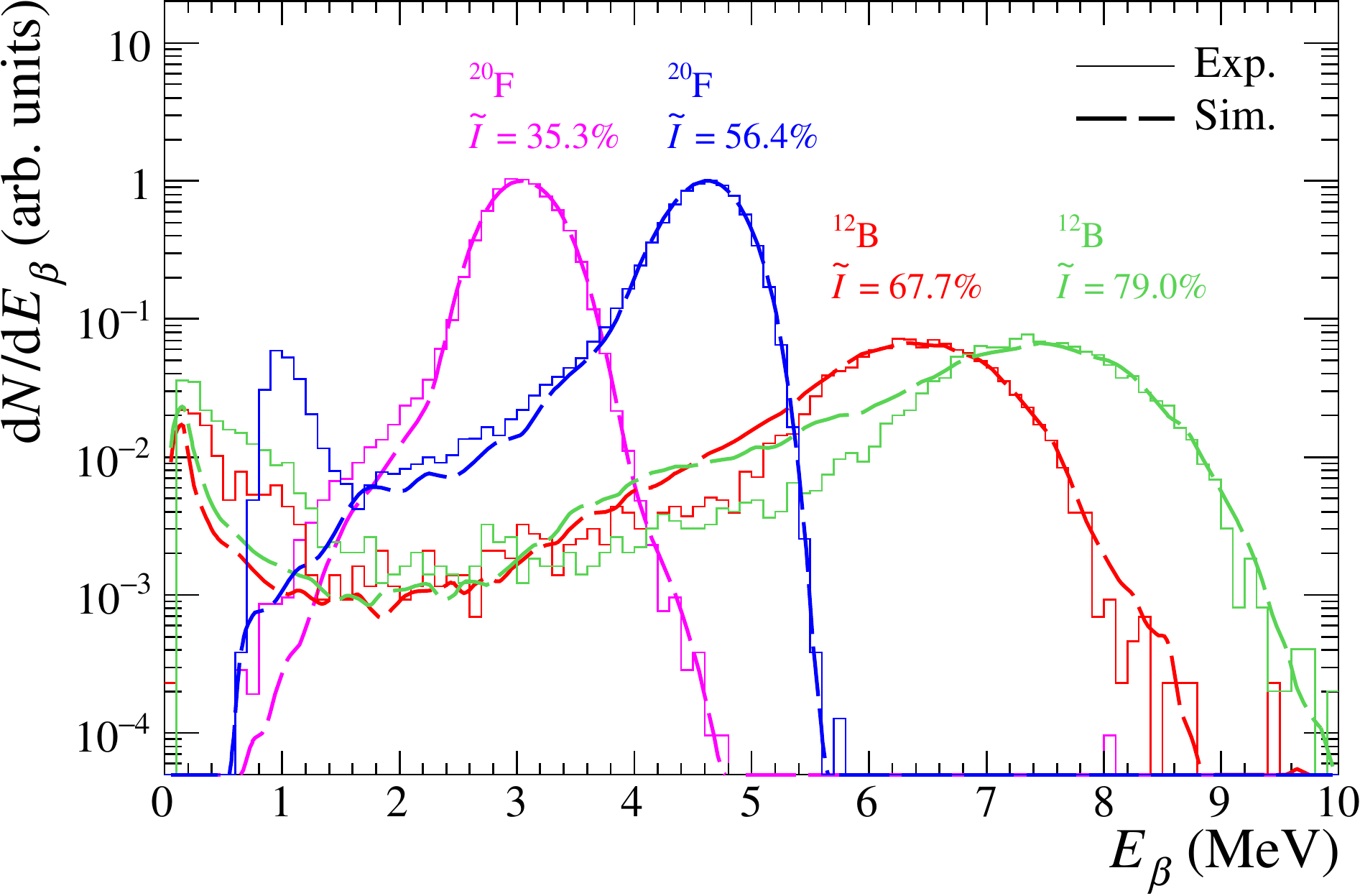}
  \caption{\label{fig:specShape} Comparison of experimental and
    simulated energy spectra obtained at 35.3\%, 56.4\% ($^{20}$F),
    67.7\% and 79.0\% ($^{12}$B) of the maximum electrical current.
    The $^{20}$F data have been subject to both veto and front cut,
    while the $^{12}$B data have been subject only to the veto cut.}
\end{figure}

By only allowing electrons within a relatively narrow energy band 
to reach the detector, the magnetic transporter effectively ``carves out'' 
a slice of the $\beta$ spectrum. %
This is demonstrated in Fig.~\ref{fig:specShape}, which shows energy spectra 
measured at three different magnetic-field strengths. %
The central energy selected by the magnetic transporter is approximately given by 
$\widetilde{E}_{\beta} \simeq 7.72 \tilde{I} + 3.01 \tilde{I}^2$~MeV, where 
$\tilde{I}$ is the electrical current expressed as a fraction of the maximum 
current provided by the power supply (700~A).
The spectra obtained at $\tilde{I}=35.3\%$ and 56.4\% show a central slice and 
the upper end of the allowed $\beta$ spectrum of $^{20}$F, respectively. %
The spectra obtained at 67.0\% and 79.0\% show slices of the $\beta$ spectrum of $^{12}$B, 
which has an end-point energy of 13.37~MeV. 
%
%
%
In all cases, the main peak is well reproduced by the GEANT4 simulation. 
Deviations occur in the low-energy tails, especially for the spectra obtained at 
the higher current settings, but these deviations are not important for the 
present analysis. %
The $^{20}$F spectra have been cleaned by requiring that no 
coincident signal is recorded in the veto detector (veto cut) and that 
the energy deposited in the front detector is between 0.65--1.60~MeV 
(front cut). %
The $^{12}$B spectra have also been subject to 
the veto cut, but the front cut could not be applied to these spectra 
because the $^{12}$B measurements were performed with the 
two-channel $\beta$ detector. %

\begin{figure}[htb]
  \includegraphics[width=\linewidth]{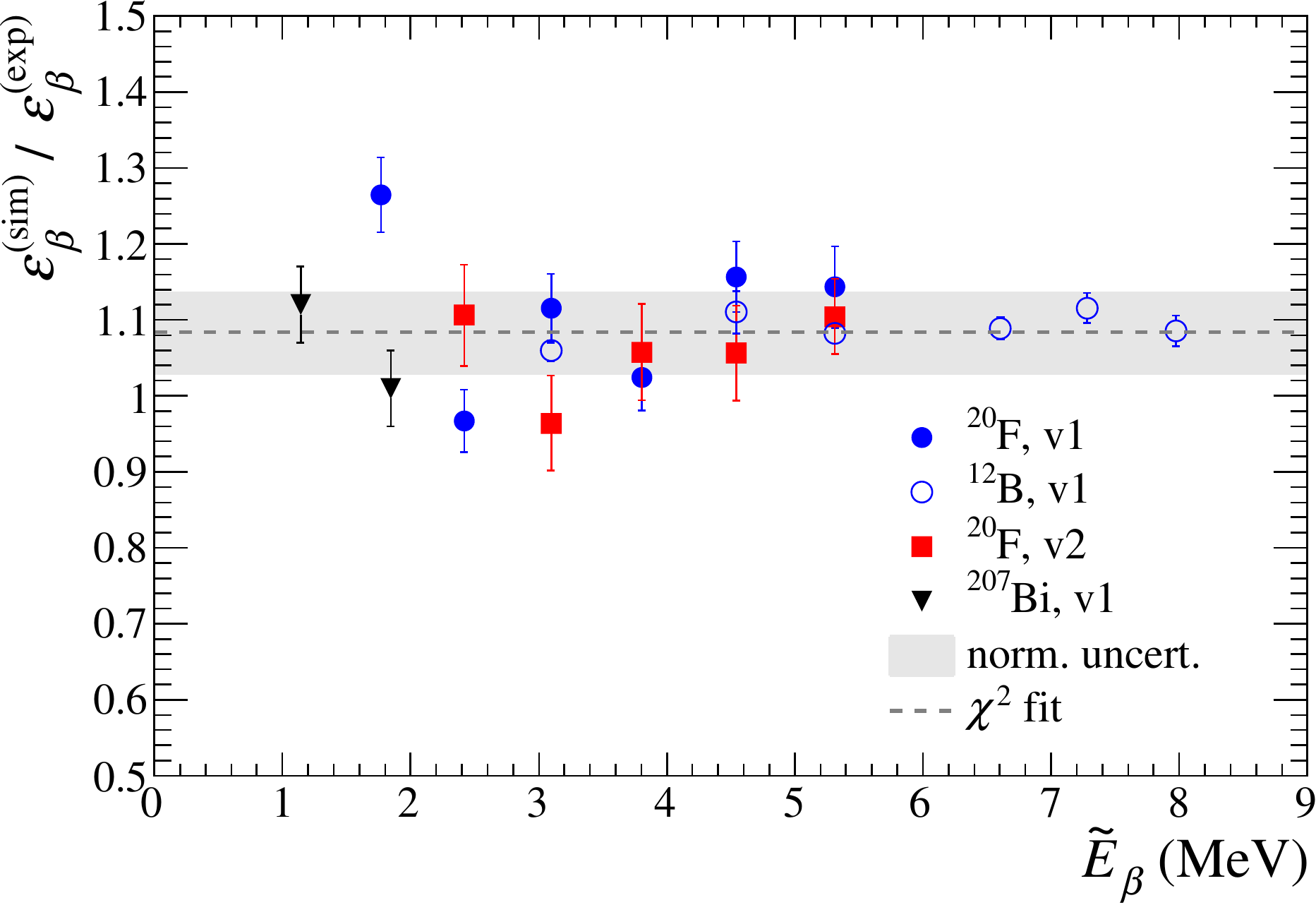}
  \caption{\label{fig:transmEff} Ratio of simulated and experimental
    transmission efficiencies,
    $\varepsilon_{\beta}^{\mathrm{(sim)}} /
    \varepsilon_{\beta}^{\mathrm{(exp)}}$, versus the central energy
    selected by the magnetic transporter, $\widetilde{E}_{\beta}$.
    The data are labeled by the isotope ($^{12}$B, $^{20}$F,
    $^{207}$Bi) and the detector (v1, v2) used for the measurement.
    The gray band indicates the systematic uncertainty on the
    normalization of the $^{20}$F data.  The dashed line was obtained
    from a $\chi^2$ fit to the data in which the normalization of the
    $^{12}$B data was allowed to vary freely (see text for details).}
\end{figure}

The $^{20}$F and $^{12}$B data (only a subset of which are shown in 
Fig.~\ref{fig:specShape}) and data obtained with a calibrated 
$^{207}$Bi source, have been used to validate the absolute accuracy 
of the GEANT4 simulations all the way up to $\widetilde{E}_{\beta} = 8.0$~MeV. %
As shown in Fig.~\ref{fig:transmEff}, the simulated and experimental $\beta$ 
transmission efficiencies exhibit reasonable agreement across 
the full energy range, with the simulation overestimating the
transmission efficiency by 8\% on average. The transmission efficiency 
is determined as $\varepsilon_{\beta} = N_{\beta} N_{\gamma}^{-1} \varepsilon_{\gamma}$, 
where $N_{\beta}$ is the number 
of counts in the full-energy peak in the uncleaned $\beta$ spectrum 
({\it i.e.}\ before application of the veto and front cuts), $N_{\gamma}$ 
is the number of 1.63-MeV $\gamma$ rays, and $\varepsilon_{\gamma}$ 
is the $\gamma$-ray detection efficiency, cf.\ Sec.~\ref{sec:abs-norm}. 
We note that the overall normalization of the $^{12}$B data points could 
not be established experimentally due to the lack of a sufficiently intense 
$\gamma$-ray line. However, by monitoring the $\beta$ count rate we were 
able to establish that the implantation rate was constant throughout the 
measurements, implying that the data points share the same overall normalization; 
its value was determined via a $\chi^2$ fit to the data. 
The 8\% overestimation may partly or entirely be attributed to the uncertainty 
on the $\gamma$-ray detection efficiency, which causes a 5\% 
uncertainty on the normalization of the experimental transmission efficiency, cf.\ Sec.~\ref{sec:abs-norm}.

\begin{figure}[h]
  \includegraphics[width=\linewidth]{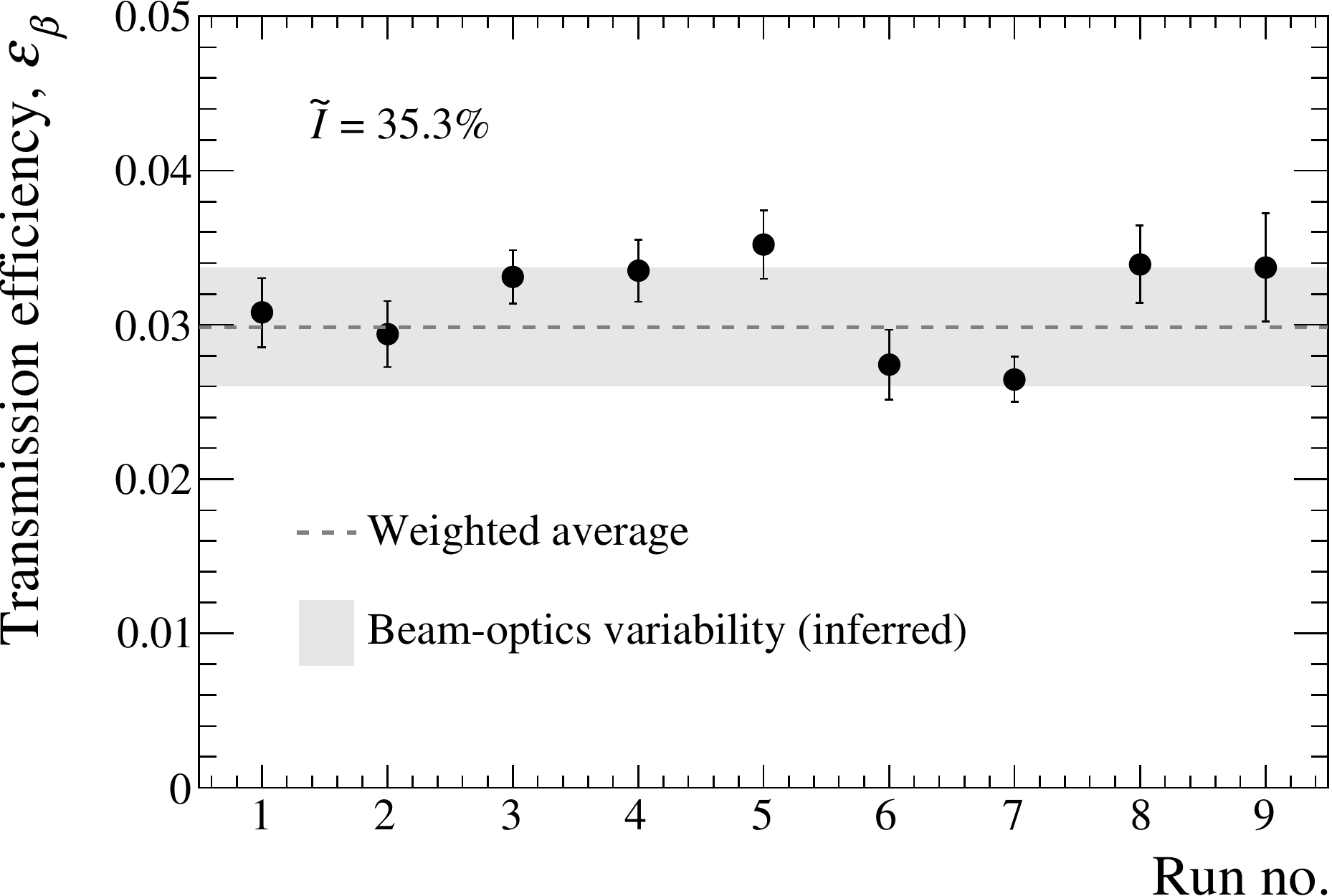}
  \caption{\label{fig:transmSpread} $\beta$ transmission efficiency
    obtained in nine separate runs performed at a magnetic-field
    strength of $\tilde{I} = 35.3\%$ at different times during the
    experiment.  The gray band indicates the spread attributed to
    variations in beam optics.}
\end{figure}

On the other hand, the large scatter in the experimental data points observed 
in Fig.~\ref{fig:transmEff} may be attributed to temporal variations in 
beam optics, which affect the source geometry and hence the transmission 
efficiency. The occurrence of such temporal variations is evident in 
Fig.~\ref{fig:transmSpread}, which shows the transmission efficiency 
obtained in nine separate runs performed at the same magnetic-field strength 
($\tilde{I} = 35.3\%$) at different times during the experiment.
The temporal variations amount to a 13\% spread in transmission efficiency, 
which we include as a systematic uncertainty on the final result.

\begin{figure}[htb]
  \includegraphics[width=\linewidth]{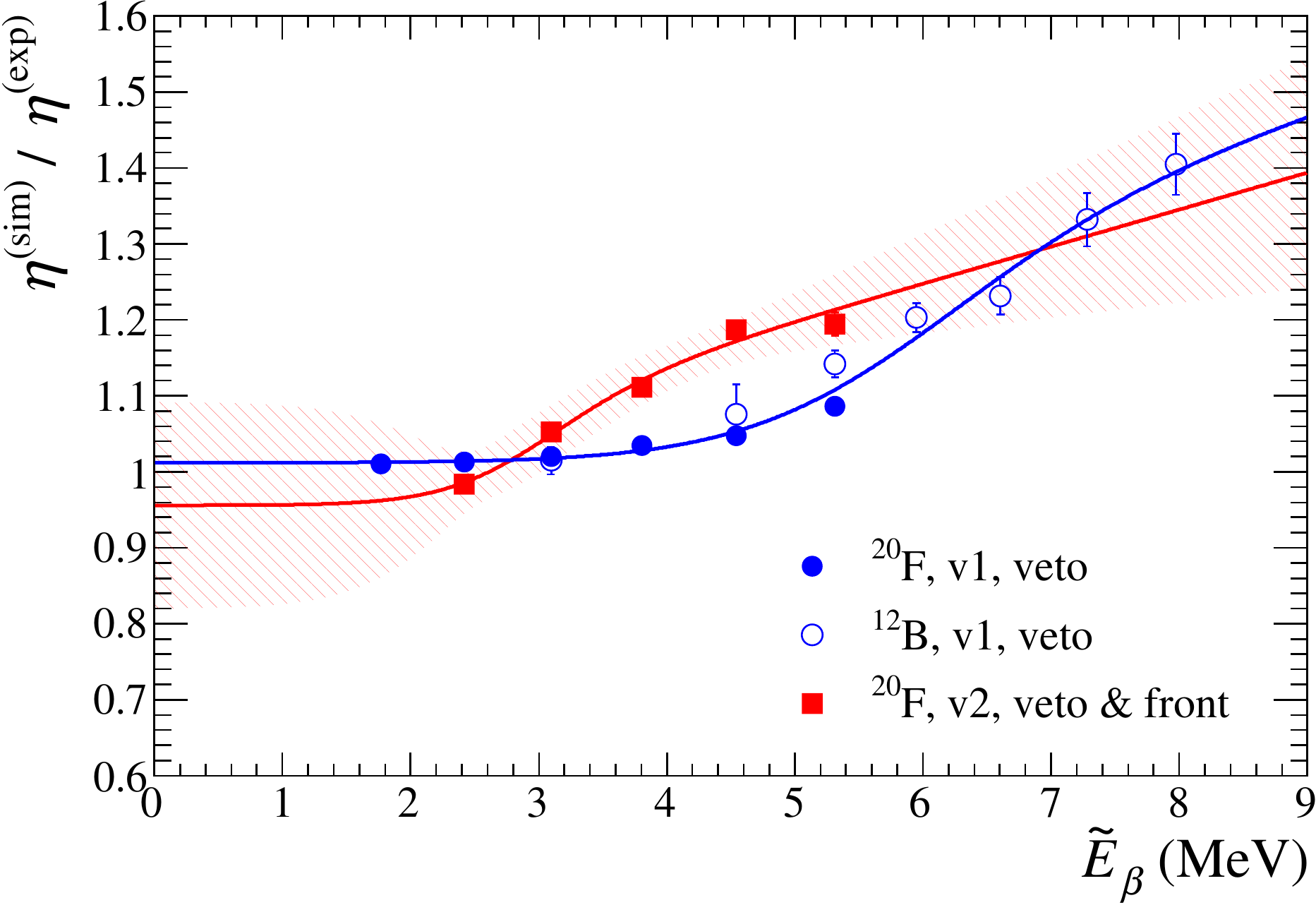}
  \caption{\label{fig:cutAcc} Ratio of simulated and experimental cut
    acceptances, $\eta^{\mathrm{(sim)}} / \eta^{\mathrm{(exp)}}$,
    versus the central energy selected by the magnetic transporter,
    $\widetilde{E}_{\beta}$.  The data are labeled by the isotope
    ($^{12}$B, $^{20}$F, $^{207}$Bi) and the detector (v1, v2) used
    for the measurement, and the cuts applied in the data analysis
    (veto \& front).  The solid lines and the hatched area show trend
    lines and the estimated uncertainty, respectively.  }
\end{figure}

Finally, we examine the cut acceptance, $\eta$, defined as the fraction 
of counts in the full-energy peak that survive the veto and front cuts. 
As shown in Fig.~\ref{fig:cutAcc}, the simulation tends to 
overestimate the cut acceptance, partly due to the presence of cross-talk 
between the inner and the outer detectors, but also due to inaccuracies 
in the modeling of the stopping process in the detector volumes. 
The factor by which the simulation overestimates the cut 
acceptance is small for $\widetilde{E}_{\beta}$, but grows 
with increasing energy reaching 1.25(9) at 
$\widetilde{E}_{\beta}\sim 6.0$~MeV. 

In Table~\ref{tbl:sys} we summarize the sources of systematic uncertainty 
affecting the normalization of the $\beta$ spectrum.  
In each case we give the estimated correction factor 
to the normalization of the GEANT4 simulation at 
$\widetilde{E}_{\beta}\sim 6.0$~MeV and the estimated
uncertainty. 
\begin{table}[h]%
  \centering
  \caption{Sources of systematic uncertainty in the normalization of the experimental $\beta$ spectrum. In each case, we give the correction factor 
  by which the spectrum has been multiplied (second column) and the uncertainty on this factor (third column).
  The total uncertainty was obtained by adding the individual contributions in quadrature.}
  \label{tbl:sys}
  \begin{tabular}{ccc}
    \toprule
    Source of sys.\ uncert.\  &  Corr.\ factor  &  Uncert.\ \\
    \colrule
    $\gamma$-ray detection efficiency  &  1  &  5\%  \\
    $\beta$ transmission efficiency  &  1/1.08  &  13\%  \\
    cut acceptance  &  1/1.25  &  7\%  \\
    \colrule
    total  &  1/1.35  &  16\% \\
    \botrule   
  \end{tabular}
\end{table}%

\subsection{Long-duration measurements}

\begin{figure}[htb]
  \includegraphics[width=\linewidth]{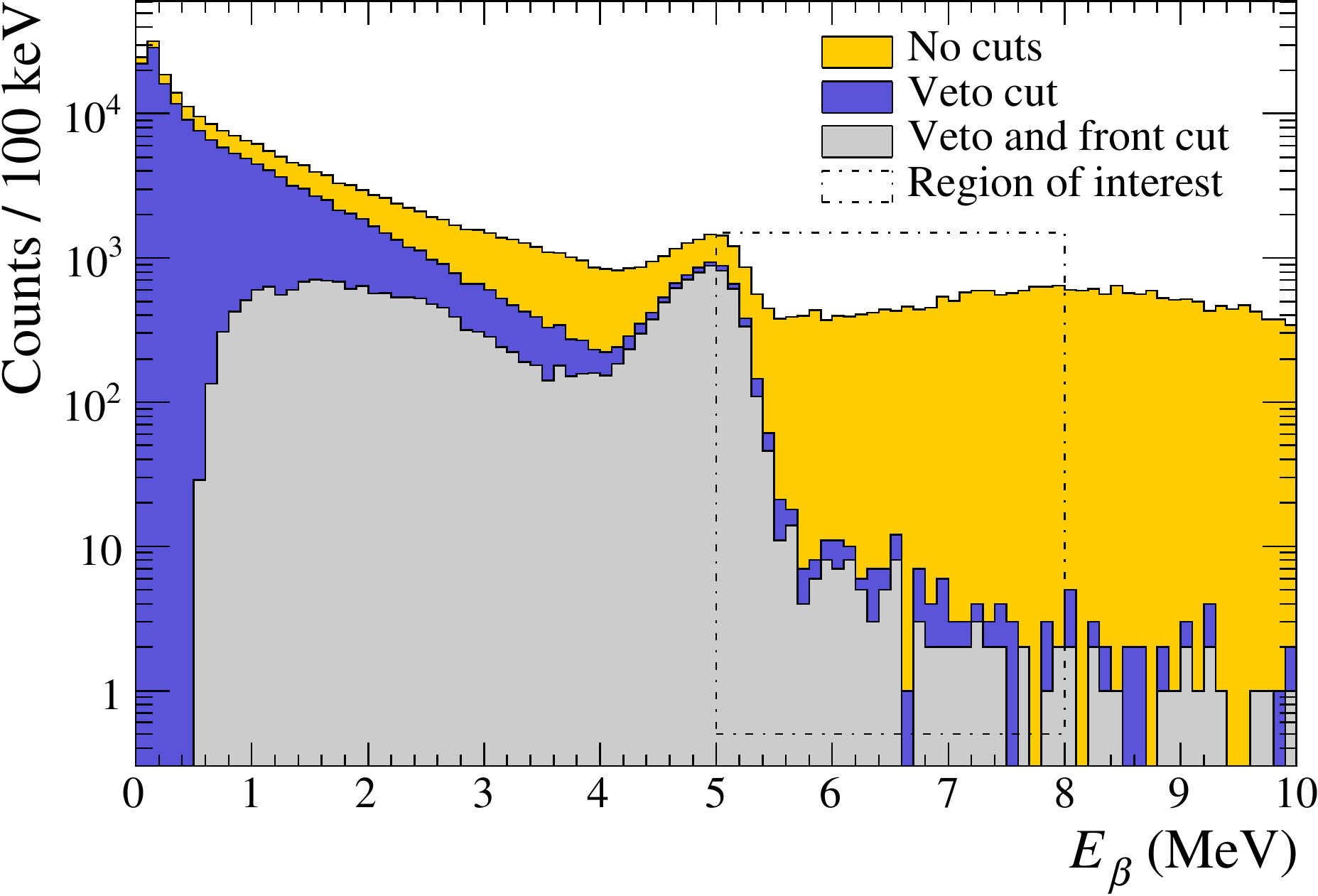}
  \caption{\label{fig:beta} $\beta$ spectrum with cuts (gray) and
    without (yellow) obtained in the inner plastic-scintillator
    detector at 67.7\% of the maximum electrical current.  The dashed
    box indicates the region used for the likelihood fits shown in
    Fig.~\ref{fig:forbidden}.}
\end{figure}

Long-duration measurements were performed at the current settings 
$\tilde{I}=67.7\%$ (67~h) and 70.5\% (38~h) to search for a signal 
in the energy region 5.4--7.0~MeV, and at 79.0\% (37~h) to demonstrate 
that any signal detected at the two lower settings did not persist 
above 7.0~MeV. %
The average $^{20}$F implantation rate for these measurements was 
11~kHz, while the $\gamma$ and $\beta$ count rates were at most a few 
tens of Hz and a few Hz, respectively, implying negligible dead time. %
Additionally, background data were collected at 67.7\% and 70.5\% for a total of 183~h. %
The $\beta$ spectrum obtained at 67.7\% is shown in 
Fig.~\ref{fig:beta}. %
The cosmic-ray background dominates the raw $\beta$ spectrum above 5.4~MeV while electrons 
from the allowed $2^+\rightarrow 2^+$ transition produce the bump centered at 5.0~MeV and the 
continuum below it. %
In the signal region the cosmic-ray background is reduced by a factor of 100 
by the veto cut. The front cut removes another factor of 3.5 resulting in an 
overall reduction factor of 350. %
On the other hand, about 2/3 of the $\beta$ particles survive the cuts, a fraction 
which would have been even larger in the absence of cross-talk between 
the inner and outer detectors. %

\subsection{Investigation of the residual background}

\begin{figure}[htb]%
  \includegraphics[width=\linewidth]{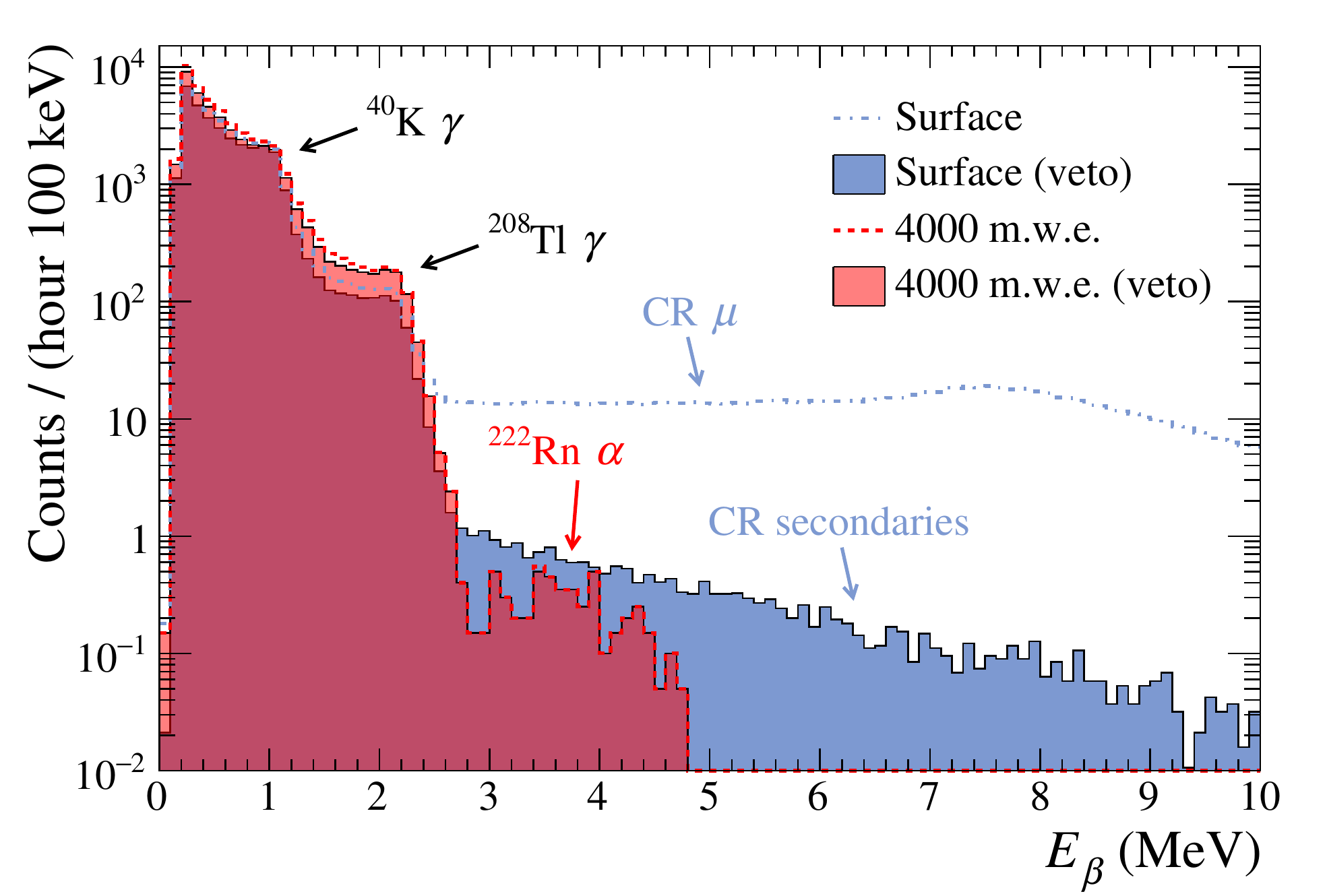}
  \caption{\label{fig:underground}The duration of the surface and
    underground measurements were 189~h and 20~h, respectively. The
    broad peaks at 1.0~MeV and 2.2~MeV are the Compton shoulders of
    the 1.46-MeV and 2.61-MeV $\gamma$-ray lines due to the naturally
    occurring radioactive isotopes $^{40}$K and $^{206}$Tl ($^{232}$Th
    decay chain).  The broad peak centered at 3.8~MeV is produced by
    the 5.49-MeV $\alpha$ particles from the decay of $^{222}$Rn,
    which lose a minimum of 0.66~MeV in the 6-$\mu$m Mylar window of
    the $\beta$ detector before entering the scintillation volume,
    plus any energy loss in the air \emph{en route} to the detector.}
\end{figure}

Below 3~MeV, $\gamma$ rays, chiefly from the decays of $^{40}$K and $^{208}$Tl, are 
the main source of background, while cosmic-ray muons dominate above this energy, 
resulting in a background rate of 150 counts/h in the signal region (5.8--6.8~MeV). 
Measurements performed at different times of the year and 
different locations within the laboratory verified that this rate was very nearly 
constant to within a few percent. %
The residual background rate in the signal region after application of the veto cut was 
1--2 counts/h. The energy dependence of the residual background is markedly different 
from the energy dependence of the raw background, indicating a different 
physical origin. %
In order to further characterize the residual background, a background measurement 
was performed at Callio Lab in the Pyh{\"a}salmi mine in Pyh{\"a}j{\"a}rvi, Finland, 
at the depth of 1430 meters (4100 m.w.e.) where the cosmic-ray muon 
flux is greatly suppressed~\cite{jalas2017}. %
In Fig.~\ref{fig:underground} we compare the spectrum obtained underground to a 
spectrum obtained at the surface. This comparison clearly demonstrates that the 
residual background is cosmic-ray induced. 
We note that the $^{222}$Rn room background activity at Callio Lab is ten times higher than 
at the JYFL laboratory (200 vs.\ 20~Bq/m$^3$), which explains the enhanced $^{222}$Rn $\alpha$ 
peak in the underground spectrum. %
Additional measurements performed at the surface with the detector fully shielded 
on all sides by 5~cm of lead further showed that $\gamma$ rays cannot be the 
main component of the residual background. We therefore conclude that hadronic 
secondaries, and neutrons in particular, from cosmic-ray interactions in the atmosphere 
and the roof of the laboratory are the likely source of the residual background. %
Finally, we note that placing the detector inside the magnetic transporter had little or 
no effect on the residual background. %
Similarly, the magnetic field seemed to exert little or no influence 
on the residual background although changes at the level of $10\%$ or 
below cannot be excluded. %

\subsection{Absolute normalization}
\label{sec:abs-norm}

\begin{figure}[h]
  \includegraphics[width=\linewidth]{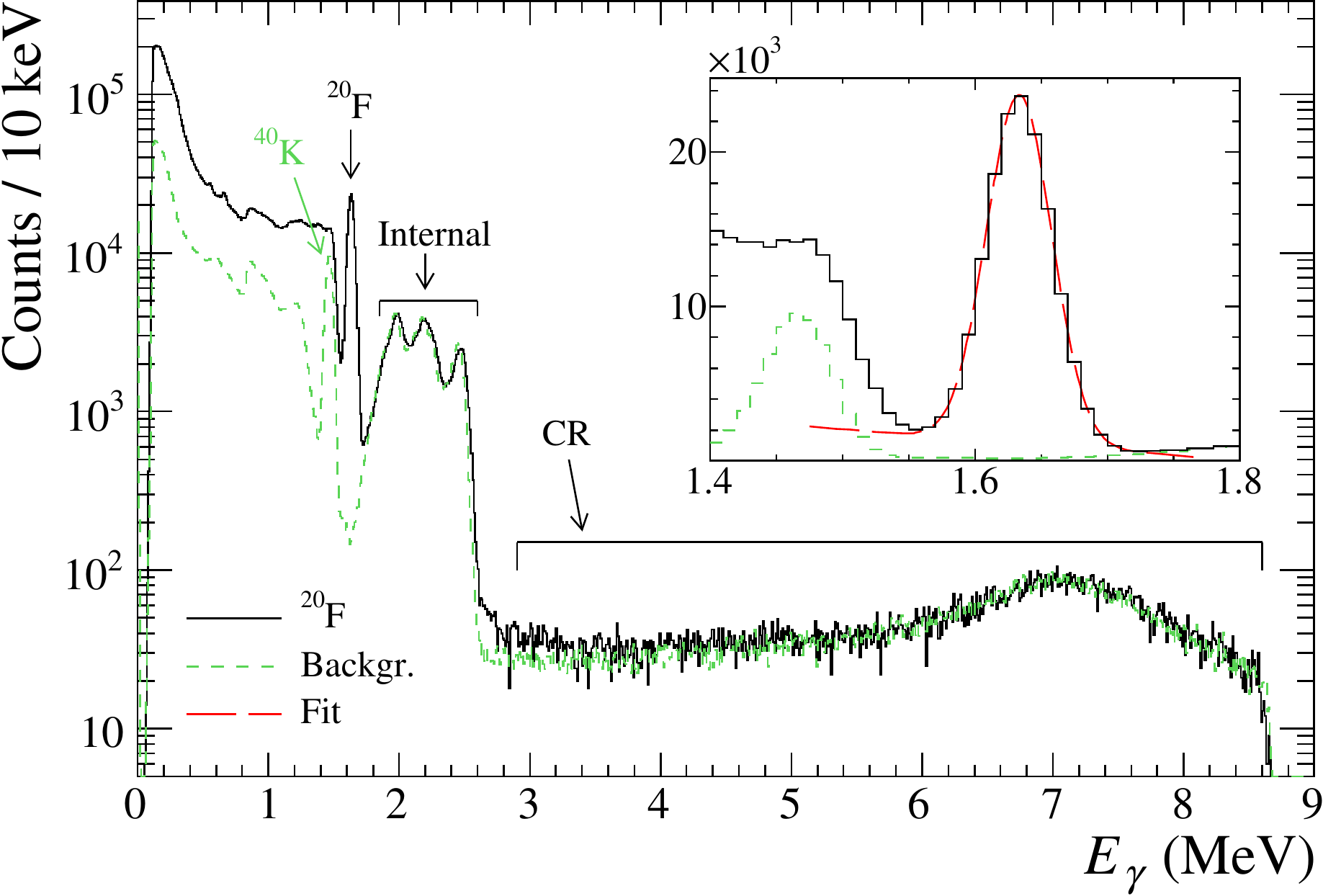}
  \caption{\label{fig:gamma} $^{20}$F $\gamma$-ray spectrum (solid,
    black line) and background $\gamma$-ray spectrum (short-dashed,
    green line) obtained in the LaBr$_3$(Ce) detector at 67.7\% of the
    maximum electrical current.  The characteristic 1.63-MeV line of
    $^{20}$F sits between the 1.46-MeV background line due to $^{40}$K
    and the peaks at 1.8--2.6~MeV due to the internal activity of the
    LaBr crystal.  Cosmic-ray muons dominate above 3~MeV. The inset
    shows a zoom-in on the 1.63-MeV line and the line-shape fit
    (long-dashed, red line).}
\end{figure}

Returning to the long-duration $^{20}$F measurements, 
we show in Fig.~\ref{fig:gamma} the $\gamma$ spectrum obtained at 67.7\%. %
The well-resolved 1.63-MeV line from the decay 
of $^{20}$F, which is used for absolute normalisation, is clearly visible. %
The efficiency of the LaBr$_3$(Ce) $\gamma$-ray detector at 1.63~MeV was determined 
online from the ratio of $\beta\gamma$ coincidences and $\beta$ singles events, yielding the value $\varepsilon_{\gamma} = 0.59(3)\times 10^{-4}$. 
This online calibration was confirmed by an offline 
calibration made using radioactive sources of $^{137}$Cs, $^{207}$Bi, $^{152}$Eu 
and $^{60}$Co of known activities, which exhibit $\gamma$-ray lines with known relative 
intensities covering the energy range 0.3--1.8~MeV. %

\subsection{Detection of the forbidden transition}

\begin{figure}[htb]
  \includegraphics[width=\linewidth]{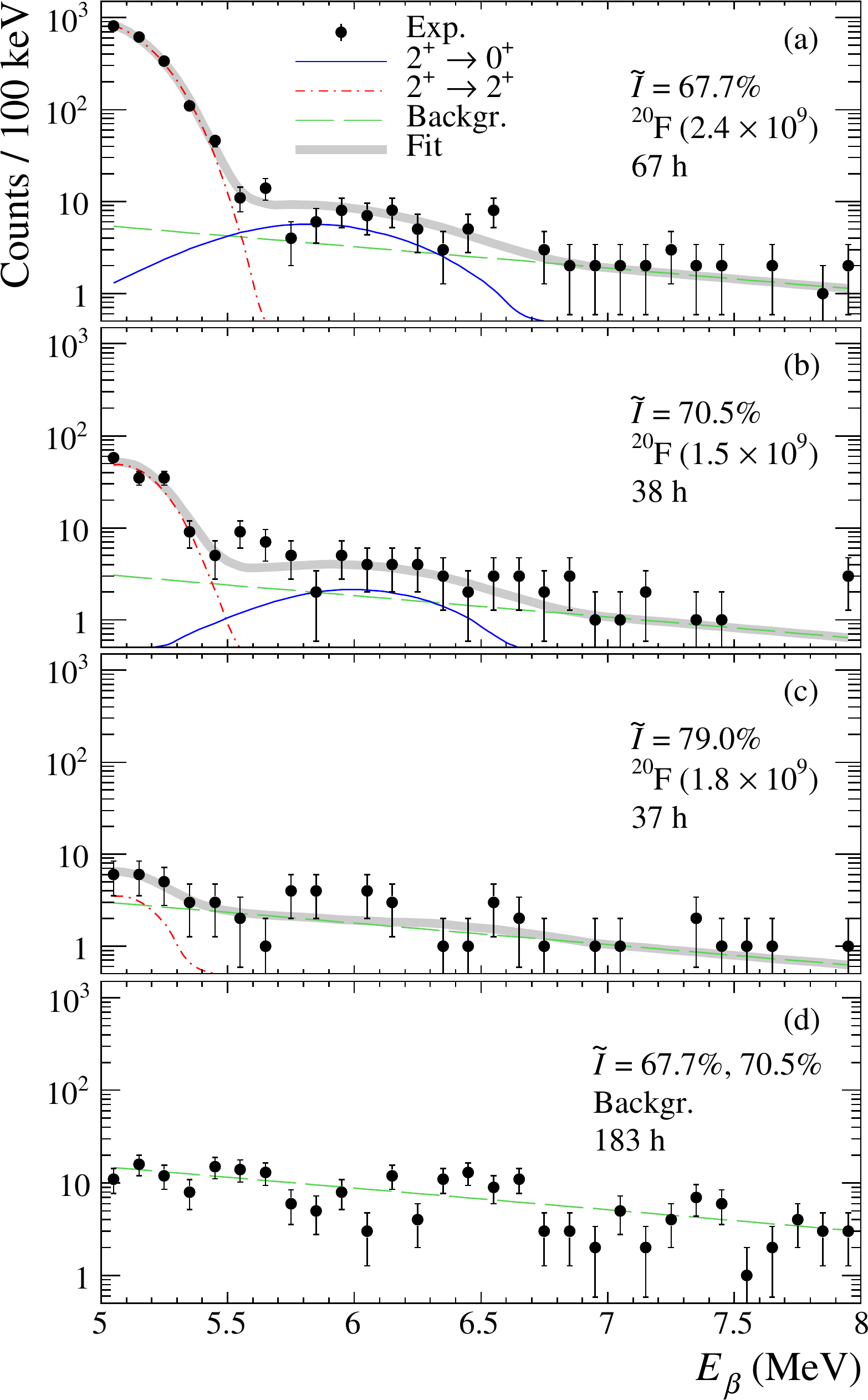}
  \caption{\label{fig:forbidden} Likelihood fits to the experimental
    data in the energy region 5.0--8.0~MeV obtained with beam (a--c)
    and without (d), at current settings focused on the region of
    interest (67.7\% and 70.5\%) and immediately above it
    (79.0\%). The contribution of the second-forbidden, ground-state
    transition in the $\beta$ decay of $^{20}$F is shown by the thin,
    solid (blue) curve.  }
\end{figure}

\begin{figure}[h]
  \includegraphics[width=\linewidth]{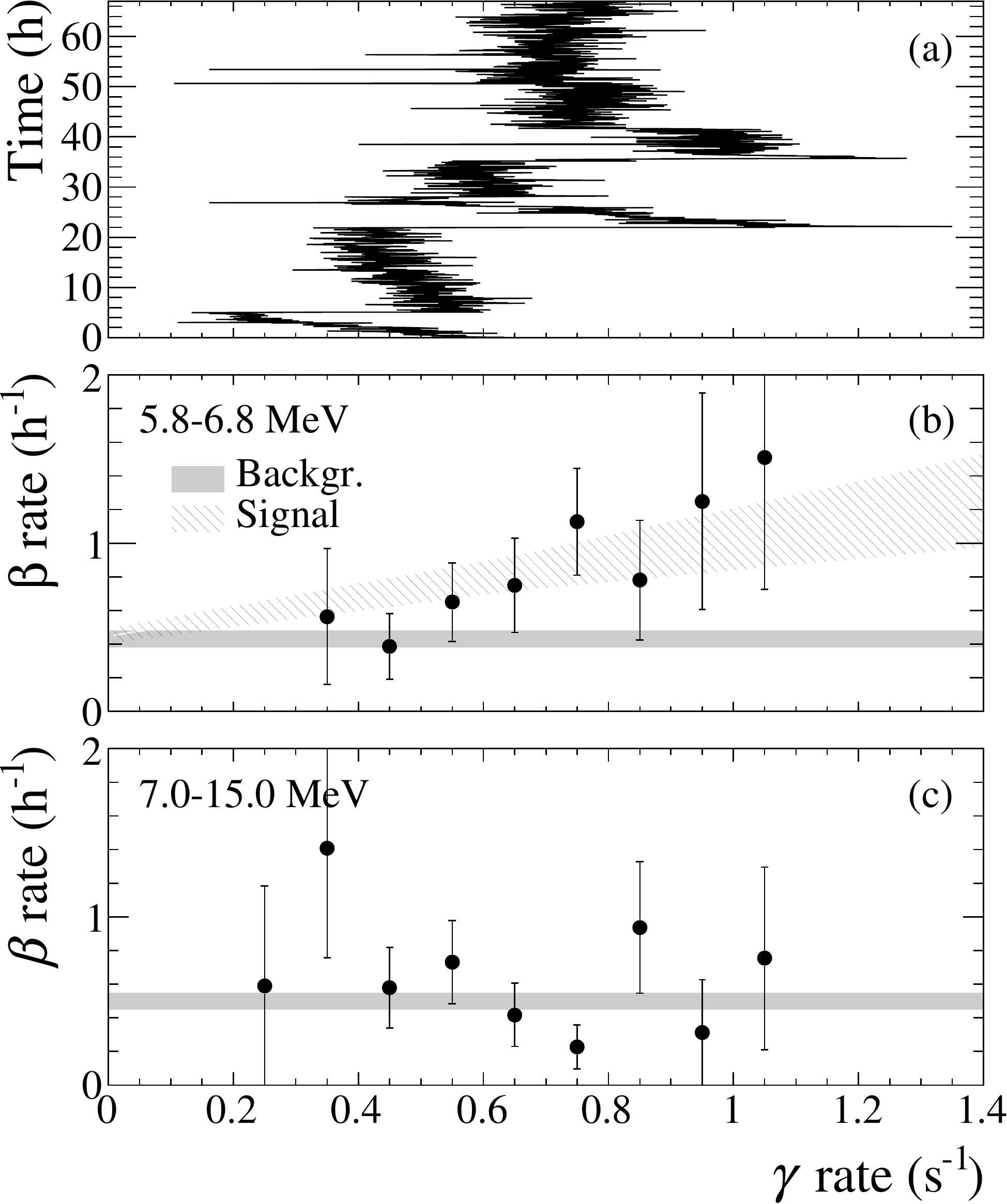}
  \caption{\label{fig:signal_yield_corr} (a) Temporal variations in
    $\gamma$-ray count rate during 67 hours of measurement at
    $\tilde{I}=67.7\%$. (b) Correlation between $\gamma$-ray count
    rate and $\beta$ count rate in the energy region
    $E_{\beta} = 5.8$--6.8~MeV.  (c) Same as (b) but for the energy
    region $E_{\beta} > 7.0$~MeV.}
\end{figure}

Fig.~\ref{fig:forbidden} shows the cleaned $\beta$ spectra obtained in the 
long-duration measurements, zoomed in on the region of interest. %
The spectra obtained at 67.7\% and 70.5\% reveal a clear excess of counts in the signal region 
when compared to the background spectrum. %
For example, the spectrum at 67.7\% has $55\pm 7$ counts between 5.8--6.8~MeV while the 
background spectrum, downscaled to account for the longer measurement time, 
only has $30\pm 3$ counts in the same region. %
Equally important, no excess of counts is observed above the signal region in the 
data obtained at 79.0\%. %
Based on the measurements performed at lower current settings (Fig.~\ref{fig:beta}) we 
can exclude $\beta\gamma$ summing and $\beta\beta$ pile-up as possible explanations. %
Furthermore, the analysis of the temporal distribution of the counts between 5.8--6.8~MeV 
shown in Fig.~\ref{fig:signal_yield_corr} reveals a clear correlation with the $^{20}$F implantation rate, which varied by 
more than a factor of two during the experiment, while the temporal distribution 
of the counts above 7.0~MeV shows no such correlation. %
Thus, the observed signal is consistent with being due to the hitherto unobserved, 
second-forbidden, ground-state transition in the $\beta$ decay of $^{20}$F. %

\section{Branching ratio}
\label{sec:branching-ratio}

The previous section has provided evidence that the second forbidden
transition connecting the ground states of $^{20}$F and $^{20}$Ne has
been measured. In order to convert the signal observed between 
5.8--6.8~MeV into a branching ratio and hence determine the magnitude 
of the matrix element, we must estimate the unobserved part of 
the forbidden $\beta$ spectrum below 5.8~MeV, where the decay is 
dominated by the allowed transition. We do so guided by shell-model 
calculations as described in the following.

\subsection{Shell-model calculations}
\label{sec:shell-model-calc-1}

For the calculation of the second-forbidden $\beta$-decay rate and $\beta$
spectrum we follow the formalism of Behrens and
B\"uhring~\cite{Behrens.Buehring:1971,*Behrens.Buehring:1982}. This
formalism accounts for the distortion of the electron wave function
due to the nuclear charge, which leads to the appearance of additional
matrix elements when compared to the formalism of 
Walecka~\cite{Walecka:1975,Donnelly.Peccei:1979} used in Ref.~\cite{Suzuki.Zha.ea:2019}.

The second-forbidden $\beta$-decay rate between the ground states of
$^{20}$F and $^{20}$Ne can be expressed as
\begin{equation*}
	    \lambda^{\beta^-} = \frac{\ln{2}}{K}\int_1^{q} C(w)wp(q-w)^2F(Z,w)dw
\end{equation*}
where $w = (E_{\beta}+m_e c^2)/m_e c^2$ is the total electron 
energy in units of $m_ec^2$, $p=\sqrt{w^2-1}$ is the electron 
momentum in units of $m_e c$, and $q = (M_p c^2 - M_d c^2)/(m_e c^2)$ 
is the energy difference between the initial and final nuclear 
state, $M_p$ and $M_d$ being the nuclear mass of the parent 
and daughter nucleus. 
The constant $K=6144\pm2$~s has been determined from superallowed 
Fermi transitions~\cite{Hardy.Towner:2009}. 
$F(Z,w)$ is the Fermi function, which arises due to the Coloumb 
interaction between the electron and the daughter nucleus with 
atomic number $Z$.
Finally, $C(w)$ is the shape factor, which depends on the matrix 
elements of the transition. For the $2^+\rightarrow0^+$ second-forbidden 
non-unique transition the shape factor has the form
\begin{equation*}
	C(w) = a_0+\frac{a_{-1}}{w}+a_1 w+a_2 w^2+a_3 w^3+ a_4 w^4
\end{equation*}
with the coefficients $a_n$ given by combinations of seven 
matrix elements~\cite{Behrens.Buehring:1971,*Behrens.Buehring:1982,Sadler.Behrens:1993}.

\begin{table}[htb]%
  \centering
  \caption{Matrix elements determining the shape factor of the
    second-forbidden transition between the ground states of $^{20}$F
    and $^{20}$Ne. The second column shows the values obtained from a
    shell-model (SM) calculation with the USDB interaction. The third
    column shows the values obtained from a shell-model calculation in
    which the matrix elements have been constrained based on the
    conserved vector current (CVC) theory and the E2 strength of
    the decay of the isobaric analog state of $^{20}$F to the ground
    state of $^{20}$Ne. The quenched matrix elements can be obtained
    by multiplying the axial values by a factor $1/g_A$. (See text for additional information.) \label{tbl:ffc}}
  \begin{ruledtabular}
    \begin{tabular}{lcc}
       Matrix element & SM & SM+CVC+E2 \\ \hline
      \VF{211}            &   0          & $-$0.0118\footnotemark[1]\\
      \VF{220}            &   0.252    & 0.184\footnotemark[1] \\
      \VF{220}$(1,1,1,1)$ &   0.301    & 0.220\footnotemark[1] \\
      \VF{220}$(2,1,1,1)$ &   0.287    & 0.210\footnotemark[1] \\
      \AF{221}            &  $-$0.122  & $-$0.122 \\
      \AF{221}$(1,1,1,1)$ &  $-$0.142  & $-$0.142 \\
      \AF{221}$(2,1,1,1)$ &  $-$0.135  & $-$0.135 \\
    \end{tabular}
  \end{ruledtabular}
  \footnotetext[1]{Matrix elements constrained from experimental data}
\end{table}

We have performed shell-model calculations in the $sd$-shell valence
space using the USDB interaction~\cite{brown2006} and the code
NUSHELLX~\cite{nushellx2014}. For the evaluation of the many-body
matrix elements we use the single-particle matrix element expressions
provided in Ref.~\cite{Behrens.Buehring:1971} modified to account for
the fact that our shell-model calculations follow the
Condon-Shortley~\cite{Condon.Shortley:1951} phase convention instead
of the Biedenharn-Rose phase
convention~\cite{Biedenharn.Rose:1953}. The resulting matrix elements
are shown in Table~\ref{tbl:ffc}. Our calculations use
harmonic-oscillator single-particle wave functions with a radial
parameter of $b=1.86$~fm and a uniform charge radius of
$R=3.88$~fm. These values reproduce the root-mean-square radius of
$^{20}$Ne determined from X-ray spectroscopy of muonic
atoms~\cite{Fricke.Bernhardt.ea:1995}.  Using Wood-Saxon wave
functions instead of harmonic-oscillator wave functions, we obtain
very similar matrix elements.

One limitation of our $0\hbar\omega$ sd-shell calculations is that the
relativistic matrix element $\VF{211}$ is identically zero for
harmonic-oscillator wave functions. This is not the case for
Wood-Saxon wave functions, but the value obtained
($\VF{211} = -0.004$) is too small to affect the results. Extending
the model space to include multi-$\hbar\omega$ excitations is beyond
the goals of the present publication and hence we follow a different
approach to determine the $\VF{211}$ matrix element. Following
Ref.~\cite{Behrens.Buehring:1971} the conserved vector current (CVC)
theory provides a relationship between \VF{211} and \VF{220},
\begin{equation}
  \label{eq:cvc}
  \VF{211} = -\frac{1}{\sqrt{10}} \left(\frac{E_\gamma R}{\hbar c}\right)
  \VF{220} \; , 
\end{equation}
where $E_\gamma = 10.273$~MeV is the excitation energy of the isobaric
analog state of the ground state of $^{20}$F in
$^{20}$Ne~\cite{Tilley.Cheves.ea:1998}.  The CVC relation
(\ref{eq:cvc}) is expected to hold for the ``exact'' matrix elements,
but may break for matrix elements computed in a restricted model space
using the impulse approximation as in our case. To quantify this
effect we further relate the magnitude of the matrix element \VF{220}
to the experimentally determined E2 strength of the decay of the
10.273~MeV state to the ground state of $^{20}$Ne assuming isospin
symmetry.
\begin{equation}
  \label{eq:vf220vsE2}
  |\VF{220}| = \frac{1}{R^2}\left(\frac{8\pi
      B(E2)}{e^2}\right)^{1/2}.  
\end{equation}
Adopting the experimental strength of $B(E2)=0.306(84)$~$e^2$~fm$^4$, we
obtain $|\VF{220}| = 0.184(25)$, while the sign of the matrix element is
determined based on the shell-model results. The matrix elements
$\VF{220}(1,1,1,1)$ and $\VF{220}(2,1,1,1)$ contain a slightly
different radial dependence than the factor of $r^2$ appearing in
$\VF{220}$. We assume that the ratios $\VF{220}(1,1,1,1)/\VF{220}$ and
$\VF{220}(2,1,1,1)/\VF{220}$ are well described by the shell-model
calculations. The full set of matrix elements obtained in this way are
listed in the column labeled ``SM+CVC+E2'' in
Table~\ref{tbl:ffc}. 

\begin{figure}[htb]
  \centering
  \includegraphics[width=\linewidth]{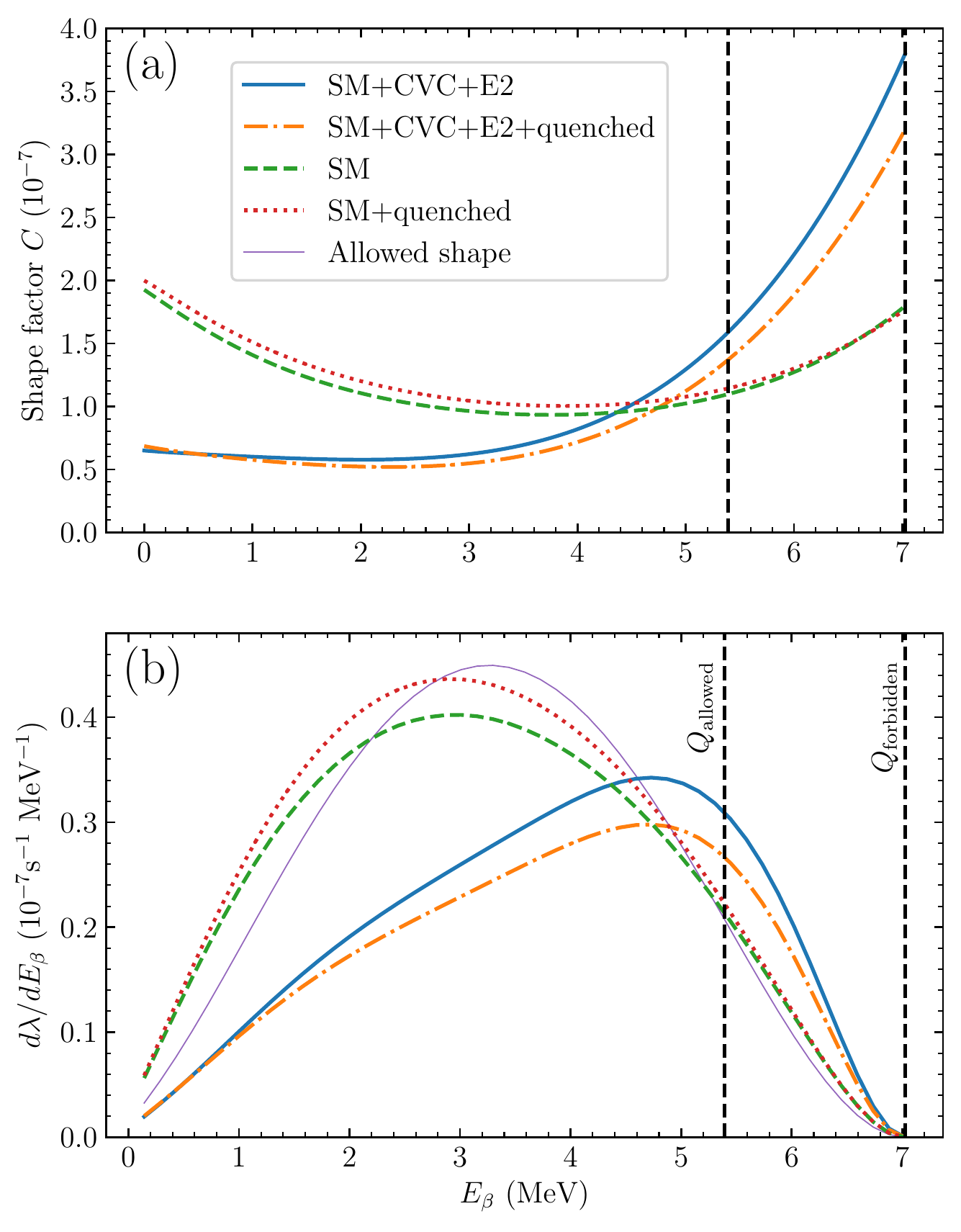}
  \caption{Theoretical shape factor (a) and $\beta$ spectrum (b) for the
    second forbidden transition as a function of the electron kinetic
    energy for the different cases discussed in text and
    table~\ref{tbl:ffc}.  The dashed vertical lines indicate the
    end-point energies of the allowed ($Q_{\text{allowed}}$) and
    forbidden ($Q_{\text{forbidden}}$) transition, respectively. The
    thin violet line shows the allowed spectral shape arbitrarily
    normalized to the ``SM'' case. \label{fig:theoshape}}
\end{figure}

The axial matrix elements $\AF{221}$,
$\AF{221}(1,1,1,1)$, and $\AF{221}(2,1,1,1)$ could be
affected by the quenching of the axial coupling constant observed in
Gamow-Teller decays, see e.g. Ref.~\cite{Towner:1987}. However, previous
studies have not shown conclusive evidence that such a quenching is
also present in non-unique second-forbidden
transitions~\cite{Warburton:1992,Martinez-Pinedo.Vogel:1998,suhonen2017}. Hence,
we will consider in the following two different cases using either the
bare value of $g_A= -1.27$ or the quenched value $g_A = -1.0$. The
numbers in table~\ref{tbl:ffc} have been obtained for $g_A=-1.27$.

In Fig.~\ref{fig:theoshape} we show the shape factor and $\beta$
spectrum of the second forbidden transition as a function of the
electron kinetic energy.  The theoretical $\log ft$ values for the
different cases are shown in the last column of
Table~\ref{tbl:br}. Looking at the shape factor and $\beta$ spectrum
one can notice important differences between the purely theoretical
results, labeled ``SM'', and those constrained by experimental
information, labeled ``SM+CVC+E2''.  In the former case the shape is
slightly distorted towards low energies compared with the allowed
shape while in the latter case it is quite distorted towards high
energies. This distortion originates from the important contribution
of the relativistic matrix element \VF{211} to the $w, w^2$ and $w^3$
terms that dominate the shape factor at high energies. For energies
around the allowed $Q$-value, $w = 5.391/0.511 + 1 = 11.55$, the shape
factor can be expressed as,


\begin{widetext}
\begin{equation}
  \label{eq:4}
  C(11.55)  =  1.3\times 10^{-6} \Bigl[(\AF{221})^2 + 58.75 \AF{221}
                 \VF{211} 
           + 1231 (\VF{211})^2 + 1.405 \AF{221} \VF{220}
            + 81.08 
           \VF{211}  \VF{220} + 1.777  (\VF{220})^2       \Bigr]  
\end{equation}
where the ratios $\VAF{KLs}(k_e,1,1,1)/\VAF{KLs}$ have been obtained 
from the shell-model calculations. One can see the important
role of the relativistic matrix element \VF{211} in determining the
shape factor at high energies. For the two limiting cases
considered above we have,

\begin{equation}
  \label{eq:2}
  C(11.55) = 1.3\times 10^{-6} \left\{
    \begin{array}{cl}
      (\AF{221})^2 + 1.405 \AF{221} \VF{220} +  1.777  (\VF{220})^2, &
      \VF{211} = 0 \\[3mm]
      (\AF{221})^2 - 2.346 \AF{221} \VF{220} +  1.619  (\VF{220})^2, &
      \VF{211}\ \text{from Eq.~\eqref{eq:cvc}}
    \end{array}
\right.
\end{equation}
\end{widetext}
From the relative signs of the vector and axial matrix elements given
in Table~\ref{tbl:ffc}, their interference is destructive for the first
case and constructive in the second case. Furthermore, quenching leads
to an small enhancement of the shape factor in the first case and a
larger reduction in the second case. 

As an additional validation of our theoretical approach, we have also computed 
the shape factor using a more general formalism that includes next-to-leading-order 
nuclear-matrix elements~\cite{haaranen2017}. We find that these additional 
matrix elements have negligible influence on the shape factor of the forbidden transition.

In the next section, we combine
the theoretical shape factors with the experimental $\beta$ spectrum
to determine the branching ratio of the forbidden transition.

\subsection{Likelihood fits to the experimental $\bm{\beta}$ spectrum}
\label{sec:shell-model-calc-1}

\begin{table*}[tb]%
  \centering
  \caption{Effect of the adopted shape factor on the fit quality and
    the inferred branching ratio and $\log ft$ value of the forbidden
    transition. We use the following notation: SM: Shell-model
    calculation; CVC+E2: Non-zero relativistic matrix element
    inferred from the CVC relation using the experimental E2
    strength. The theoretically preferred shape factor is indicated in
    bold font.  The fits with and without the forbidden transition
    have $N=112$ and $113$ degrees of freedom, respectively. For the
    branching ratio ($b_{\beta}$) we give the statistical fit uncertainty
    first, followed by the systematic experimental uncertainty; these
    are added in quadrature, including also the (significantly
    smaller) uncertainties on the end-point energy and the total
    half-life, to obtain the final uncertainty on the $\log ft$
    value.}
  \label{tbl:br}
  \begin{ruledtabular}
  \begin{tabular}{ccccccc|c}
    Forbidden transition  &  Shape  & $g_A$ & $\chi^2/N$  &  $p$-value  &  $b_{\beta}$ $(\times 10^{-5})$  &  $\log ft$  &  $\log ft$ (theory) \\
    \colrule
    \textbf{yes}  &  \textbf{SM+CVC+E2} & $\bm{-1.27}$  &  \textbf{1.193}  &  \textbf{0.080}  &  $\textbf{0.41(8)(7)}$    &  $\textbf{10.89(11)}$  &  $\textbf{10.86}$ \\
    yes  &  SM+CVC+E2  & $-1.0$   & 1.190  &  0.083  &  $0.43(8)(7)$    &  $10.88(11)$  & $10.91$ \\
    yes  &  SM              & $-1.27$  & 1.190  &  0.083  &  $0.90(17)(14)$  &  $10.55(11)$  & $10.76$ \\
    yes  &  SM              & $-1.0$   & 1.189  &  0.083  &  $0.95(18)(15)$  &  $10.53(11)$  & $10.73$ \\
    yes  &  allowed         & -        & 1.192  &  0.081  &  $1.10(21)(18)$  &  $10.46(11)$  &  - \\
    \colrule
    no   &  -   & - & 1.518  &  0.00032  &  0  &  -  &  -  \\ 
  \end{tabular}
\end{ruledtabular}
\end{table*}%

\begin{table}[htb]%
  \centering
  \caption{Quality of the likelihood fits to the spectra in
    Fig.~\ref{fig:forbidden} performed with the theoretically 
    preferred forbidden shape factor (``yes'') and assuming 
    no contribution from the forbidden transition (``no'').}
  \label{tbl:fits}
  \begin{ruledtabular}
  \begin{tabular}{cccc|ccc}

    \multirow{3}{*}{Panel}  &  \multicolumn{6}{c}{Forbidden transition} \\\cline{2-7}
      &  \multicolumn{3}{c}{yes}  &  \multicolumn{3}{c}{no} \\\cline{2-7}
      &  $N$  &  $\chi^2/N$  &  $p$-value  &  $N$  &  $\chi^2/N$  &  $p$-value \\
    \colrule
    (a)  &  24  &  1.39  &  0.098  &  25  &  1.97  &  0.0027 \\
    (b)  &  24  &  1.35  &  0.12   &  25  &  1.79  &  0.0087 \\
    (c)  &  24  &  1.08  &  0.35   &  25  &  1.05  &  0.39 \\
    (d)  &  28  &  1.50  &  0.044  &  28  &  1.82  &  0.0049 \\
    \colrule
    all  &  112  &  1.19  &  0.080  &  113  &  1.52  &  0.00032 \\

  \end{tabular}
\end{ruledtabular}
\end{table}%

In order to determine the branching ratio, we perform a likelihood fit
to the experimental $\beta$ spectrum between 5.0--8.0~MeV, in which we 
allow the normalisation of the simulated spectra of the allowed 
and forbidden transitions to vary freely, while the background is modelled 
by a simple exponential function with two free parameters. While the
normalisation of the allowed transition is, in principle, fixed, in
practice it is necessary to allow the normalisation to vary because
the GEANT4 simulation becomes inaccurate in the low-energy tail of the
trasmission window. We also allow for a small ($<50$~keV) constant
energy shift to account for inaccuracies in the energy calibration.
%

We perform such a likelihood fit for each of the four forbidden shape 
factors shown in Fig.~\ref{fig:theoshape}. For reference, we also perform 
fits assuming a forbidden shape factor of unity ({\it i.e.}\ allowed shape) 
and assuming no contribution from the forbidden transition. %
The results of these fits are summarized in Table~\ref{tbl:fits}. %
Apart from the fit that ignores the contribution of the forbidden transition, 
all fits have practically identical fit qualities, implying that the shape factor 
is essentially unconstrained by the experimental data. %
As a result, the branching ratios differ substantially, ranging from 
$\sim 0.4\times 10^{-5}$ to $\sim 1.0\times 10^{-5}$, with the smaller value 
being favored by the theoretical arguments given in Sec.~\ref{sec:shell-model-calc-1}. %
We note that our result is consistent with the existing 
upper limit of $\sim 10^{-5}$, which was obtained 
assuming an allowed shape~\cite{calaprice78}. %

When comparing the theoretical and experimental $\log ft$ values we
find that the theoretical $\log ft$ values constrained by experimental data, 
labeled ``SM+CVC+E2'', are consistent with experimental $\log ft$ values, while the 
purely theoretical $\log ft$ values, labeled ``SM'', overestimate the half-life of the 
forbidden transition by a factor of $\sim 1.6$. We do not find major differences
between the quenched and unquenched cases. In the following, we will
adopt the shape factor given by the unquenched ``SM+CVC+E2'' model, 
shown in boldface in Table~\ref{tbl:br}, as this model is
consistent with all the available experimental data including the CVC theory, 
the strength of the analog E2 decay in $^{20}$Ne, and the presently measured forbidden $\beta$ 
spectrum, and there is no compelling evidence supporting the need of quenching for second forbidden 
transitions~\cite{Warburton:1992,Martinez-Pinedo.Vogel:1998}. %

Adopting this forbidden shape factor, the simultaneous fit
to the four spectra shown in Fig.~\ref{fig:forbidden} yields a
branching ratio of $0.41(8)\times 10^{-5}$ and a goodness of fit of
$\chi^2/N = 133.6/112 = 1.193$ corresponding to an acceptable
$p$-value of $P_{\chi^2>133.6} = 0.080$. %
%
%
If, on the other hand, we fix the branching ratio to zero, the goodness of fit 
worsens to $\chi^2/N =  171.5/113 = 1.52$ corresponding to a $p$-value of 
only $P_{\chi^2>171.5} = 0.00032$, providing clear evidence for a positive 
observation. %

\begin{figure}[htb]
  \includegraphics[width=\linewidth]{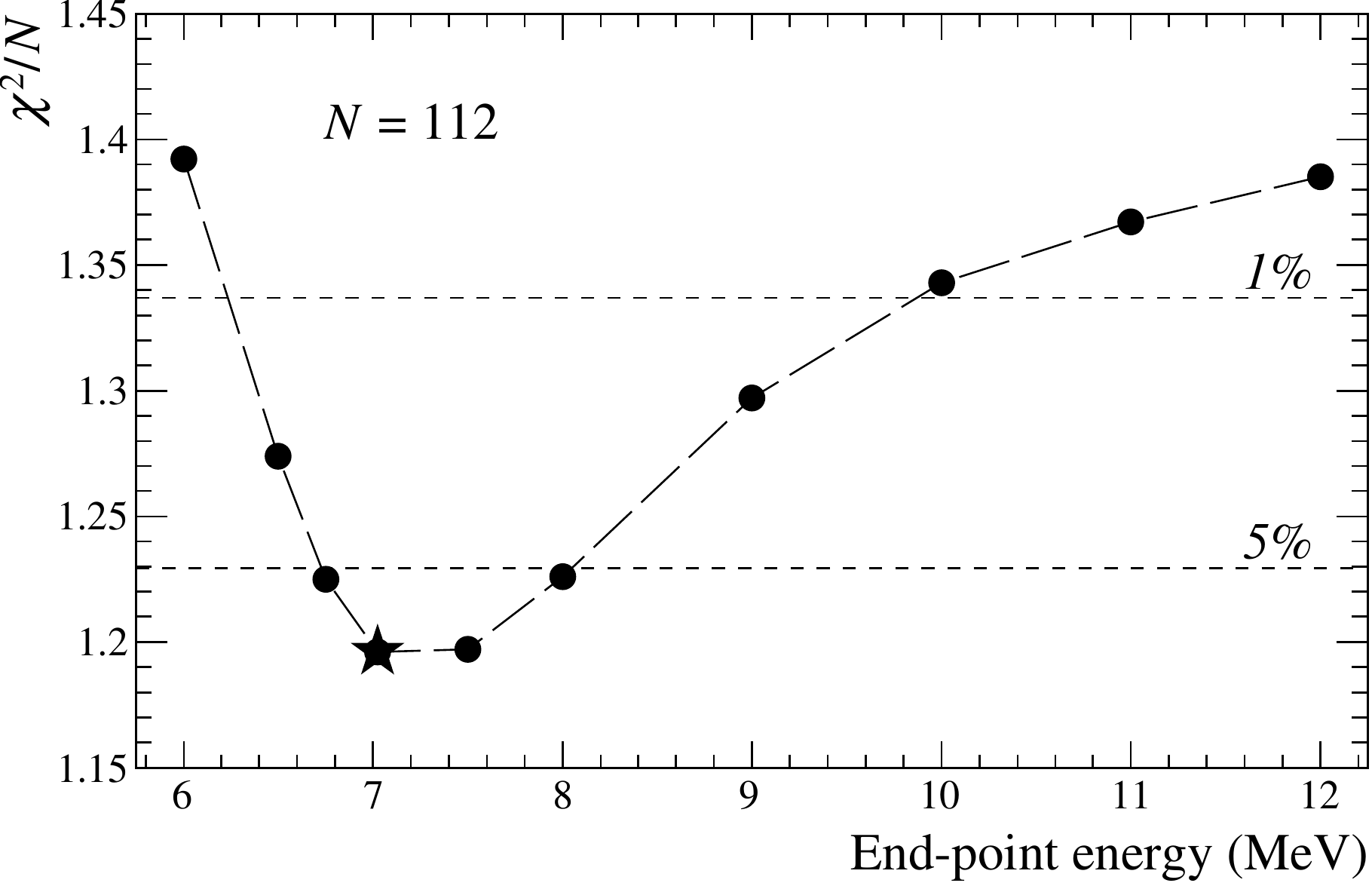}
  \caption{\label{fig:chi2} Dependence of the goodness of fit on the
    assumed end-point energy of the forbidden transition. The 5\% and
    1\% significance levels are shown by the dashed horizontal
    lines. The actual end-point energy of $7.025$~MeV is indicated by
    the star.}
\end{figure}

In Fig.~\ref{fig:chi2} we show the dependence of the goodness of fit
on the assumed end-point energy of the forbidden transition.
The best fit is obtained by adopting an end-point energy very close to 
the actual value of $7.025$~MeV. %
The 95\% confidence interval is determined to be $[6.74;\, 8.00]$ making it 
highly unlikely that an unknown $\beta^-$-unstable beam contaminant should be 
the cause of the observed signal. %
This is in accordance with expectations as $^{20}$F is the only $\beta$ emitter 
with mass 20 produced by the $^{19}{\rm F}(d,p)$ reaction at 6~MeV. %
Measurements performed on neighboring masses and on mass 40 were used to rule 
out the possibility that the signal was due to a $\beta$ emitter with a mass different 
from 20, transmitted to the setup through the tails of the acceptance window of the 
dipole magnet or as doubly-charged ions. %

Taking into account the uncertainties related to the normalisation of the $\beta$ spectrum 
discussed above, our result for the branching ratio of the forbidden transition is 
$b_{\beta} = \left[ 0.41\pm 0.08\textrm{(stat)}\pm 0.07\textrm{(sys)} \right] \times 10^{-5}$. 
This translates into $\log ft = 10.89(11)$, 
where the statistical and systematical uncertainty have been added in quadrature.

\section{Electron capture rate}\label{sec:ecrate}

The astrophysical importance of the forbidden transition 
was first pointed out in Ref.~\cite{pinedo14}, where it was argued 
that the inverse $0^+\rightarrow 2^+$ transition could enhance the rate of electron capture 
on $^{20}$Ne in dense and hot astrophysical environments, thereby 
affecting the final evolution of stars that develop degenerate 
cores of oxygen and neon. Ref.~\cite{pinedo14} also provided an estimate 
of the electron-capture rate based on the previous upper limit on the 
branching ratio of the forbidden transition~\cite{calaprice78} 
assuming an allowed shape. In the following, we generalize the calculation 
of the electron-capture rate to account for the forbidden shape.

At the high densities and temperatures of a degenerate oxygen-neon
stellar core the nuclei are fully ionized and the electrons form a
relativistic and degenerate Fermi gas. The energy of the electrons
follow the Fermi-Dirac distribution with chemical potential $\mu_e$
related to $\rho Y_e$, where $\rho$ denotes the matter density and
$Y_e$ the electron fraction. The electron capture rate via the
forbidden transition is given by,
\begin{equation}
\lambda^{EC}=\frac{\ln{2}}{K}\int_q^{\infty}C(w)wp(w-q)^2F(Z,w)S_e(w,\mu_e)dw \label{eq:ecrate}  
\end{equation}
where $q$ is the positive $Q$-value of the transition in units of the
electron mass, $S_e(w,\mu_e)$ the Fermi-Dirac distribution, and $F(Z,w)$ is the
Fermi function, where $Z$ is the charge number of the capturing nucleus. 
Screening corrections have been included in the calculation of the rate following 
Refs.~\cite{Juodagalvis.Langanke.ea:2010,pinedo14}.

\begin{figure}[htb]%
  \includegraphics[width=\linewidth]{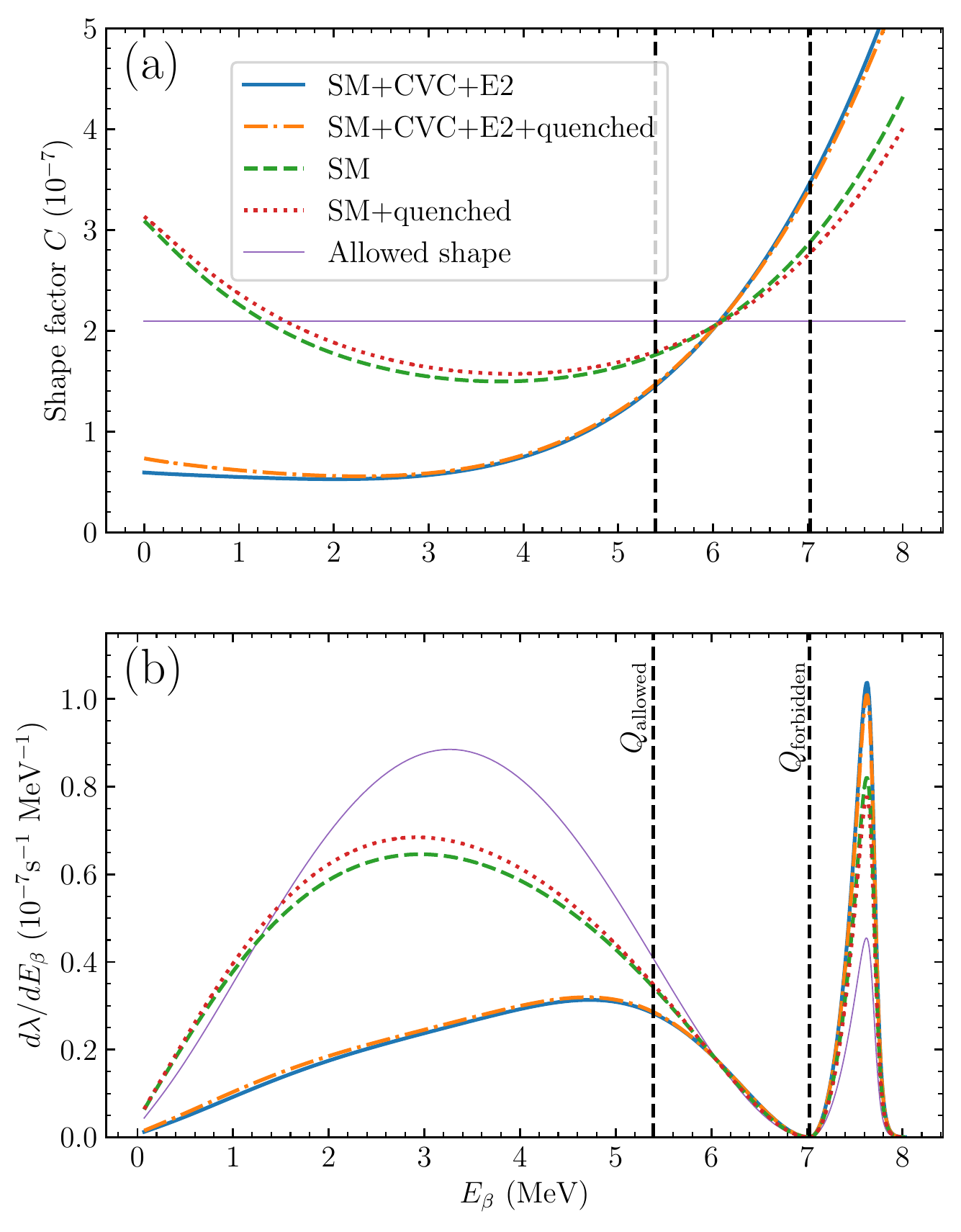}
  \caption{\label{fig:shapefactors} Normalized shape factors (a) and
    corresponding $\beta$-decay spectra (b) as a function of the
    electron kinetic energy. The two vertical lines correspond to the
    $Q$-values of the allowed and forbidden transitions,
    respectively. The experimental data constrains the spectra between
    these two values. For energies above the $Q$-value we show the
    shape factor for electron capture after dividing by a factor
    $\sqrt{5}$ in the upper panel. The lower panel shows spectra of
    captured electrons for the conditions of temperature,
    $\log_{10} T(\text{K})=8.6$~ and density
    $\log_{10} \rho Y_e (\text{g cm}^3) = 9.6$.}
\end{figure}%

We compute the shape factor of the electron-capture transition using the 
same expression as for $\beta$ decay, taking into account the different 
kinematics ($E_\nu = Q- E_e$ for $\beta^-$ decay and $E_\nu = E_e -Q$ for 
electron capture), using the same relative phases of the matrix elements 
as in Table~\ref{tbl:ffc}, and correcting for the trivial factor of $\sqrt{5}$
arising from the reversal of initial and final spins. %


Fig.~\ref{fig:shapefactors} shows the shape factors and $\beta$ spectra 
for the various cases considered including the assumption of allowed 
shape. In all cases, the shape factors have been normalized to the 
experimental $ft$ value by multiplying all matrix elements by a constant 
factor. This factor is very close to 1 for the experimentally 
constrained matrix elements, labeled ``SM+CVC+E2'', and close to 0.7 
for the purely theoretical matrix elements, labeled ``SM''. %
Above the end point of the forbidden transition
($E_{\beta} > Q_\mathrm{forbidden}$), we show 
the shape factor of electron capture
and the electron-capture rate computed for the representative
conditions of temperature $\log_{10} T(\text{K})=8.6$ and
density $\log_{10} \rho Y_e (\text{g cm}^3) = 9.6$. In the
region between the end points of the allowed and forbidden 
transitions ($Q_\mathrm{allowed} < E_{\beta} < Q_\mathrm{forbidden}$), 
the $\beta$ spectra are very similar once normalized to the experimental data. The extrapolations to lower energies ($E_{\beta} < Q_\mathrm{allowed}$) based on the theoretical shape factors differ substantially, which explains why the inferred branching ratios differ 
by more than a factor of two, cf.~Table~\ref{tbl:br}. 
Similarly, differences can be observed
at higher energies in the energy regime relevant to electron
capture ($E_{\beta} > Q_\mathrm{forbidden}$). However, due to the relative sharp cut-off of the Fermi-Dirac distribution $S_e(w,\mu_e)$ we 
do not need to extrapolate far and the
maximal difference is only $\sim 25\%$. 

\begin{figure}[h]%
\includegraphics[width=\linewidth]{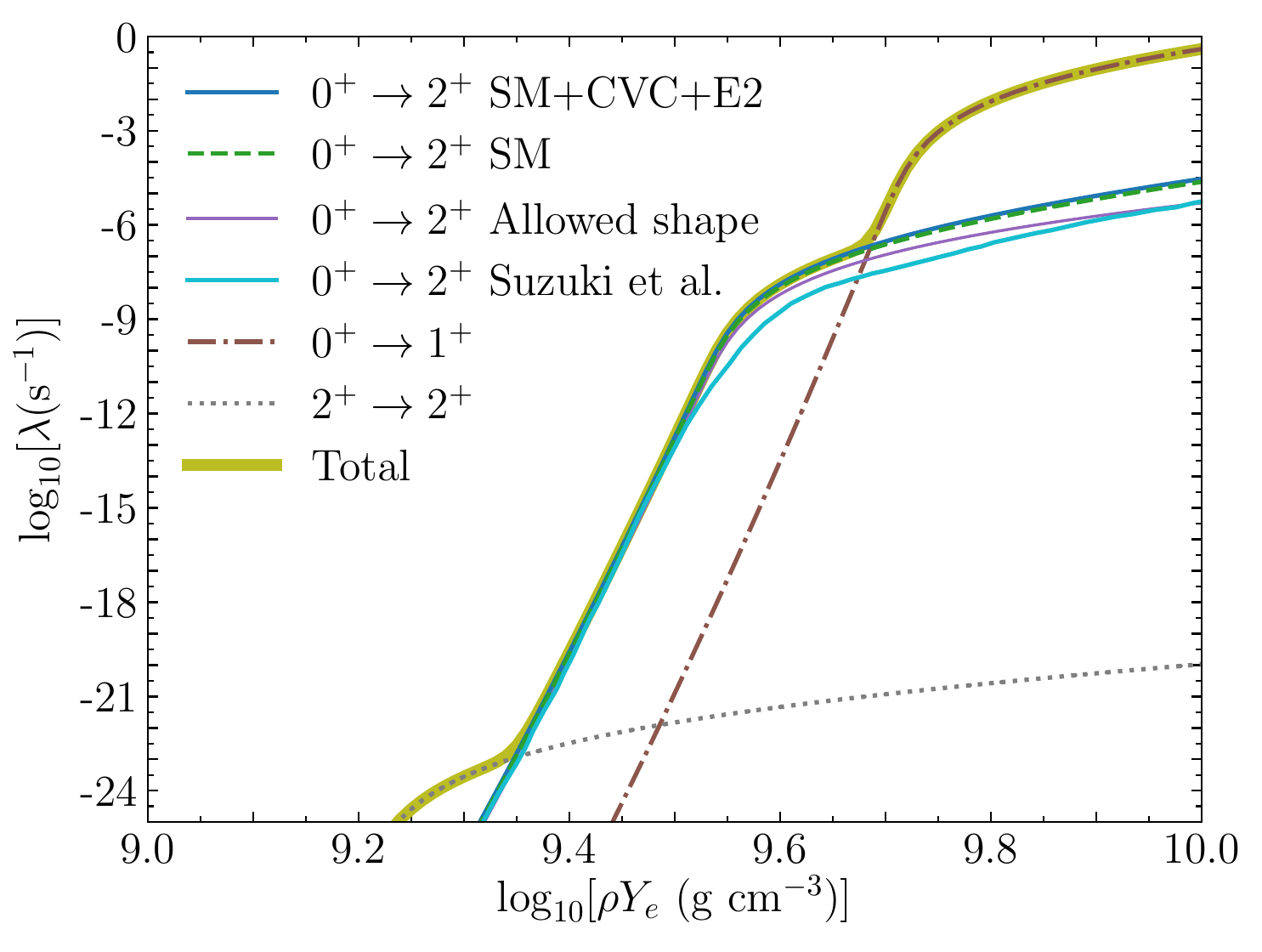}
\caption{\label{fig:ecrates}Electron capture rate on $^{20}$Ne as a
  function of density at a temperature of
  $\log_{10} T(\text{K})=8.6$. We show the contributions of 
  the forbidden transition studied in this work and two
  allowed transitions: One from the ground state of $^{20}$Ne to the $1^+$
  state in $^{20}$F at 1.056~MeV, and one from the $2^+$ state of
  $^{20}$Ne at 1.634~MeV to the ground state of $^{20}$F. The former
  dominates the rate at densities $\log_{10} \rho Y_e (\text{g cm}^3)
  \gtrsim 9.67$, whereas the latter dominates at 
  $\log_{10} \rho Y_e (\text{g cm}^3) \lesssim 9.35$. However, at such
  densities the rate is so small that it has no influence on the evolution. 
  For reference we also show the forbidden
  contribution as computed by Suzuki {\it et al.}~\cite{Suzuki.Zha.ea:2019}.}
\end{figure}%

In Fig.~\ref{fig:ecrates} we show the electron-capture rate on
$^{20}$Ne as a function of the density for a temperature 
of $\log_{10} T(\text{K})=8.6$. The chosen conditions are
representative of those reached during the evolution of 
degenerate oxygen-neon cores prior to oxygen ignition~\cite{Kirsebom.Jones.ea:2019}.
%
%
The forbidden transition is seen
to increase the electron-capture rate by several orders of 
magnitude in a critical density range compared to the case 
in which only allowed transitions
are considered. The rates computed with
the two theoretical models that account for the energy dependence 
of the forbidden transition differ by less than $25$\%. 
Such a small difference has no significant impact on the evolution of 
degenerate oxygen-neon stellar cores. %
The rate computed assuming an allowed shape is smaller than 
the two other rates. The deviation becomes larger at higher 
densities, but remains within a factor of two in the relevant 
density range where the forbidden transition dominates over the allowed transition to the $1^+$ state. %
The rate determined in Ref.~\cite{Suzuki.Zha.ea:2019} based 
on a shell-model calculation, which uses the same USDB interaction 
as in the present work, is substantially smaller than the 
present rate. The deviation reaches a factor of $\sim 10$ at 
densities of $\log_{10} \rho Y_e (\text{g cm}^3) \sim 9.6$. %
The origin of this discrepancy is not clear, but it is likely 
related to differences in the treatment of the forbidden 
transition, which lead to substantially different electron 
spectra (see their Fig.~5).

\section{Discussion}

Shell-model calculations are known to reproduce the 
strengths of second-forbidden, \emph{unique} transitions 
in the $sd$ shell within a factor of two or better~\cite{Warburton:1992,Martinez-Pinedo.Vogel:1998}. %
A similar conclusion was also reached by Ref.~\cite{Sadler.Behrens:1993}  regarding the second-forbidden, {\it non-unique} decay of $^{36}$Cl. 
%

Here, we have shown that for the second-forbidden, non-unique, $2^+\rightarrow 0^+$ transition between the ground states of $^{20}$F
and $^{20}$Ne, the accuracy is also better than a factor of two.  A purely
theoretical calculation overestimates the half-life by a factor of
$\sim 1.6$, whereas a calculation constrained by the known strength of
the analog E2 transition in $^{20}$Ne together with the CVC theory
reproduces the experimental half-life to within 10\%.
It would be of considerable interest to extend this comparison to the 
much weaker decay of $^{36}$Cl ($\log ft = 13.321(3)$~\cite{kriss2004}), which is the only other 
known second-forbidden, non-unique transition in the $sd$ shell. 
However, this is beyond the scope of the present study and is left 
for future work.

The $2^+\rightarrow 0^+$ transition in the $\beta$ decay of 
$^{20}$F was only observed in a narrow energy range near the 
end point of the $\beta$ spectrum. This, combined with limited 
statistics, a modest signal-to-background ratio, and a modest 
energy resolution, meant that the experimental data did not impose 
any useful constraints on the shape of the spectrum. Instead, the shape  was determined from a combination of theoretical calculations 
and the known strength of the analog E2 transition in $^{20}$Ne. %
This led to a branching ratio of 
$b_{\beta} = \left[ 0.41\pm 0.08\textrm{(stat)}\pm 0.07\textrm{(sys)} \right] \times 10^{-5}$ 
and a strength of $\log ft = 10.89(11)$. %
This makes the $2^+\rightarrow 0^+$ transition in the $\beta$ decay of 
$^{20}$F the second-strongest, second-forbidden, non-unique transition ever measured, with the 27 previously measured transitions having $\log ft$ values ranging from 10.6 to 14.2~\cite{singh1998}. %

\section{Conclusion}

The second-forbidden, non-unique, $2^+\rightarrow 0^+$ ground-state transition in the 
$\beta$ decay of $^{20}$F has been observed for the first time. %
The detection was made possible by the development of a dedicated 
experimental setup consisting of a Siegbahn-Sl\"atis type intermediate-image 
magnetic electron transporter combined with a plastic-scintillator telescope. %
The branching ratio was determined to be $b_{\beta} = \left[ 0.41\pm 0.08\textrm{(stat)}\pm 0.07\textrm{(sys)} \right] \times 10^{-5}$, 
implying $\log ft = 10.89(11)$, which makes this the second-strongest, second-forbidden, non-unique $\beta$ transition ever measured. %
This remarkable result is supported by our shell-model calculations, which reproduce the 
experimental strength to within better than a factor of two.
%


Owing to its large strength, the forbidden transition between the ground state of $^{20}$Ne 
and $^{20}$F enhances the astrophysical electron-capture rate on $^{20}$Ne by several 
orders of magnitude at the elevated temperatures and densities achieved in contracting 
oxygen-neon stellar cores. This has significant impact on the final evolution of 
such stars as discussed elsewhere. %
Here, we have shown that the experimental data constrain the astrophysical capture rate to within 
better than 25\%, which is fully sufficient to assess the astrophysical implications.

The experimental data did not impose any useful constraints on the shape of the forbidden 
$\beta$ spectrum, which instead was determined based on shell-model calculations and the 
experimental $B(E2)$ value of the analog transition in $^{20}$Ne. Future experiments should 
aim to provide improved constraints on the shape of the forbidden $\beta$ spectrum, although 
this will be very challenging. 

%

\begin{acknowledgments}

We are greatly indebted to the technical staff 
at the JYFL laboratory and Aarhus University who contributed 
with their time and expertise to the refurbishment 
of the spectrometer. %
We thank F.~Lyckegaard for making the BaF$_2$ targets, H.~Kettunen for preparing 
the catcher foil, and E.~Nacher for providing technical support with the 
GEANT4 simulations. %
Finally, we thank C.~Matteuzzi, M.~Anghinolfi, P.~Hansen, 
R.~Julin and T.~Kib\'edi for valuable advice and encouragement 
in the early phases of the project. %
This work has been supported by the Academy of Finland under the 
Finnish Centre of Excellence Programme 2012--2017 (Nuclear and 
Accelerator Based Physics Research at JYFL) and the Academy of 
Finland grants No.\ 275389, 284516 and 312544. %
DFS and GMP acknowledge the support of the Deutsche
Forschungsgemeinschaft (DFG, German Research Foundation) -
Projektnummer 279384907 - SFB 1245 ``Nuclei: From Fundamental
Interactions to Structure and Stars''; and the ChETEC COST action
(CA16117), funded by COST (European Cooperation in Science and
Technology).
This project has been partly supported by the Spanish Ministry MINECO 
through the grant FPA2015-64969-P and has received funding from the 
European Union's Horizon 2020 research and innovation programme under 
grant agreement No.\ 654002 (ENSAR2). %
OSK acknowledges support from the Villum Foundation through 
Project No.\ 10117. 

\end{acknowledgments}

\bibliography{refs}

\end{document}